\documentclass[openacc]{rstransa}

\titlehead{Research}

\begin{document}

\title{Alfv\'en Pulse Driven Spicule-like Jets in the Presence of Thermal Conduction and Ion-Neutral Collision in Two-Fluid Regime}

\author{
A. K. Srivastava$^{1}$, Anshika Singh$^{1}$, Balveer Singh$^{2}$, K. Murawski$^{3}$, T.V.~Zaqarashvili$^{4}$, D. Yuan$^{5}$, E. Scullion$^{6}$, Sudheer K. Mishra$^{7}$, B. N. Dwivedi$^{8}$}

\address{$^{1}$Department of Physics, Indian Institute of Technology (BHU), Varanasi 221005, India;
$^{2}$Aryabhatta Research Institute of Observational Sciences (ARIES), Manora peak, Nainital 263001, India;
$^{3}$Institute of Physics, University of Maria Curie-Sklodowska, Pl. M. Curie-Sklodowskiej, 20-0531 Lublin, Poland;
$^{4}$Institut of Physics, IGAM, University of Graz, Universitätsplatz 5, 8010, Graz, Austria;
$^{5}$Shenzhen Key Laboratory of Numerical Prediction for Space Storm, Institute of Space Science and Applied Technology, Harbin Institute of Technology, Shenzhen, Guangdong 518055, China;
$^{6}$Northumbria University, NE1 8ST Newcastle upon Tyne, UK;
$^{7}$Astronomical Observatory, Kyoto University, Sakyo, Kyoto 606-8502, Japan;
$^{8}$Rajiv Gandhi Institute of Petroleum Technology, Jais Amethi 229304, India}

\subject{Solar Physics, Sun's atmosphere, Sun's plasma}


\corres{A. K. Srivastava\\
\email{asrivastava.app@itbhu.ac.in}}

\begin{abstract}
We present the formation of quasi-periodic cool spicule-like jets in the solar atmosphere using 2.5-D numerical simulation in two-fluid regime (ions+neutrals) under the presence of thermal conduction and ion-neutral collision. The non-linear, impulsive Alfv\'enic perturbations at the top of the photosphere trigger field aligned magnetoacoustic perturbations due to ponderomotive force. The transport of energy from
Alfv\'en pulse to such vertical velocity perturbations due to ponderomotive force is considered as
an initial trigger mechanism. Thereafter, these velocity perturbations steepen into  the shocks followed by  quasi-periodic rise and fall of the cool jets transporting mass in the overlying corona.  
\end{abstract}


\begin{fmtext}




\end{fmtext}


\maketitle

\section{Introduction}

The complex magnetic field structuring and highly dynamic nature of the Sun's chromosphere permit the genesis of numerous  high speed jet-like plasma ejecta at different time and spatial scales \cite{2004Natur.430..536D,2014Sci...346A.315T,2017Sci...356.1269M,2019AnGeo..37..891S,2020ApJ...897..153G,2022MNRAS.511.4134S, 2023EPJP..138..209S,2023ApJ...945..113M}. These cool spicule-like jets are significant dynamical plasma processes that may be accompanied with various types of waves, oscillations, instabilities etc in the Sun's atmosphere \cite{2010ApJ...718.1070H,2018ApJ...856...44A,2020ApJ...889...95M,2021ApJ...923...72M,2020ApJ...897..153G,2022MNRAS.511.4134S,2023Physi...5..655S}. Such confined and cool jet like features may be one of the possible sources to carry energy and mass from the inner layers (i.e., photosphere, chromosphere) of the solar atmosphere into its upper layers 
(i.e, transition region, and corona). These jet-like phenomena evoked the scientific research since long keeping the view of the advent of many space-borne and ground-based, high-resolution solar observations, as well as theoretical/numerical modelling \cite{2004Natur.430..536D,2006ApJ...647L..73H,2007ApJ...655..624D,2009ApJ...705..272R,2010ApJ...722.1644S,2011Sci...331...55D,2012Natur.486..505W,2014Sci...346A.315T, 2016A&A...589A...3S, 2019Sci...366..890S, 2022MNRAS.511.4134S}. These jets may also have impulsive origin, and they could exhibit quasi-periodic upward and  downward plasma motions along their spine in the localized solar atmosphere. There are several observational and theoretical results available, { which} frequently { describe} their origin, evolution, physical properties and associated plasma processes \cite{2006ApJ...647L..73H,2018NatAs...2..951S,2022MNRAS.511.4134S, 2023EPJP..138..209S, 2023ApJ...945..113M}. However, their exact triggering mechanisms are still unclear, thereby detailed investigation is  always an essential requirement  using observations with high resolution observations as well as extensive numerical modelling.

Recently, multi-wavelength and high spatio-temporal resolution observations taken from ground and space based observatories reveal the triggering and evolution of small-scale cool and thin spicule-like jets.
These jets can be observed { for example} in  H$_{\alpha}$, Na D3, Ca II spectra of the solar chromosphere, which reveal their double-thread structures, detailed morphology, kinematics, and property of quasi-periodical rise and fall \cite{2004Natur.430..536D,2006ApJ...647L..73H,2007ApJ...655..624D,2009ApJ...705..272R,2010ApJ...722.1644S,2014Sci...346A.315T,2018NatAs...2..951S, 2019Sci...366..890S,2022MNRAS.511.4134S, 2023ApJ...945..113M}. Although many of them allude to the similar basic structure and plasma configuration, these cool jets have a variety of nomenclature and morphology depending on where and how they are observed. The family of cool chromospheric jets show-up distinct morphological, evolutionary, and kinematical properties (e.g., type I spicule, type II spicule, network jets, macrospicules, fibrils, mottles, swirls etc). Spicules are mainly seen on the solar limb propagating with the speed of about 25 km s$^{-1}$, which is greater than supersonic speed in the chromosphere. These cool jets trace the magnetic field lines. They are referred to type I spicules. There are plenty of studies also considering such ejecta observed on the solar disk in form of RRE/RBEs, fibrils, etc.~
{ The type I spicules have a maximum height of 7-10 Mm, typical lifetime of 5-15 min, and lateral width of 500 km.\cite{2009SoPh..260...59P}}. On the other hand, the cool jets propagate with a typical high speed of $\approx$ 50-100 km $s^{-1}$, attain their maximum height of $\approx$ 4 Mm, referred as type II spicule. Its width is $\approx$ 200 km, { and} typical lifetime $\approx$ 10-15 sec. 
These spicules are also seen rapidly fading in chromospheric absorption lines \cite{2009ApJ...705..272R, 2014Sci...346A.315T}.
This phenomenon appears due to the fact that the jet's cool plasma material quickly expand or fade away by the heating process \cite{2015ApJ...806...14P,2013PASJ...65...62T}. { The life-time of the Type II spicule also depends on the observables as seen in Ca II emissions from the solar chromosphere. They disappear, but their life-time is longer at the lines formed higher up in the solar atmosphere, e.g., C II or Si IV \cite{2015ApJ...806...14P}.} Some of type II spicules are found in coronal holes but especially in plage regions exhibiting their maximum height $\approx$ 5 Mm.
The detailed physical properties of type-I and II spicules are described in \cite{2012SSRv..169..181T}.

In addition, the large-scale similar spicule-like jets are also commonly observed in the solar atmosphere. They are termed as macrospicules. They have impulsive origin and attain a maximum height of 7-70 Mm. Their typical lifetime is about 3-15 min, and vertical flow rate is $\approx$10-150 km s$^{-1}$
\cite{2011A&A...535A..58M,2014ApJ...783...11A, 2019ApJ...871..230L, 2021MNRAS.505...50G, 2023ApJ...942L..22D}. These comparatively larger spicules may be observed in the polar region of the coronal holes. More-recently, the high spatio-temporal (SJI) and spectral resolution observations taken from Interface Region Imaging Spectrograph (IRIS) reveal the cool and thin jets mainly rooted in the magnetic networks below { the} corona with a typical apparent speed $\approx$80-250 km s$^{-1}$ , and termed as network jets \cite{2014Sci...346A.315T, 2014ApJ...792L..15P,2019ApJ...873...79C}. Their maximum height is $\approx$ 10 Mm and typical lifespan is 20-80 s. 
The quiet-Sun network jets may also rotate with an apparent velocity of $\approx$50 km s$^{-1}$\cite{2014ApJ...792L..15P, 2018A&A...616A..99K}. 
In addition 
, the larger cool jets may also exhibit the tendency of helical/rotatory motion, and their spire can acquire magnetic twists \cite{2015A&A...573A.130P,2017ApJ...848...38I ,2018A&A...616A..99K}. { In particular, the large-scale cool plasma jets and surges are usually linked with flux emergence/cancellation. 
The cool counterpart of the surges are observed in the blue/red shifted absorption lines in H-$\alpha$ observations and Ca II H \& K lines \cite{2012ApJ...752...70U, 2016ApJ...822...18N, 2017ApJ...850..153N, 2020A&A...639A..22J}. 
These surges may possess initiation velocity of about 50 km s$^{-1}$, which could  increase upto a maximum of 100-300 km s$^{-1}$ \cite{2012ApJ...752...70U}. Such cool jets may attain a height of 10–200 Mm or greater \cite{2000SoPh..196...79S,2012ApJ...752...70U}. The lifetime of surges is calculated on average about 30 minutes, while they may periodically rise and fall with a period of an hour or longer \cite{2000SoPh..196...79S,1995SoPh..156..245S,2012ApJ...752...70U}.}

Various theoretical and extensive numerical models have been { performed} in recent years to understand the physical features, triggering/formation mechanism and associated plasma processes of such cool jets. The evolution of these cool jets { may also be most likely associated with} small-scale magnetic reconnection in the  solar photosphere/chromosphere. This physical process may result in an 
{ energetic explosion} in the solar chromosphere/photosphere, which may indirectly affect the localized magneto-plasma region further resulting in the { triggering of cool jet-like motions moving} upward through the magnetic field lines \cite{2011A&A...535A..95D}. The evolution and formation of the jets are determined by the height and { the} magnitudes of the reconnection whether it is occurred { near the photosphere  or in the upper chromosphere}. 
{ When the reconnection occurs near the solar photosphere,  the plasma outflows collides with the high plasma beta region and moves along the magnetic field lines. This dynamics eventually triggers a slow-mode wave further steepens  as a strong slow shock while propagating upward in the solar atmosphere. When this shock passes through the transition region, it lifts cool plasma jets. Eventually when magnetic reconnection occurs in the  upper part of the solar chromosphere, the cool plasma is accelerated in the form of jets because of the collective effect of Lorentz  force and the transverse whip-like motions of the magnetic field lines} \cite{1982SoPh...77..121S,2008ApJ...683L..83N, 2011A&A...535A..95D, 2013ApJ...770L...3K,2013PASJ...65...62T}. Another classical scenario of the formation of such cool jets is the onset of energy deposition by pressure pulses or velocity pulses at chromospheric/photospheric heights. As a result, eventually an evolution of the magnetoacoustic shock enables the { motion of the} plasma in form of the { cool jets}. In such models, the applied pulses act as a driver and create localized heating or transfer of momentum resulting into shocks and generating the jet-like features  \cite{1982SoPh...77..121S,2004Natur.430..536D,2006ApJ...647L..73H,2007ApJ...666.1277H,2010A&A...519A...8M, 2018NatAs...2..951S, 2019AnGeo..37..891S, 2022MNRAS.511.4134S}. 
{ The process generally proposed as the agents causing} the formation of cool spicular jets and other chromospheric cool plasma ejecta (e.g., cool jets, network jets, macrospicules, spicules, swirls etc) is the shock-generated model \cite{1982SoPh...77..121S,2004Natur.430..536D,2006ApJ...647L..73H,2007ApJ...666.1277H, 2011A&A...535A..58M, 2011ApJ...736....9M, 2019AnGeo..37..891S, 2022MNRAS.511.4134S, 2023EPJP..138..209S}. { Hansteen et al \cite{2006ApJ...647L..73H} reported the formation  of cool jet-like plasma ejections (e.g., dynamic fibrils, mottles etc) from the active regions driven by slow-mode upward propagating magnetoacoustic shocks. The formation of these shocks is caused by waves motion triggered by the random convective flows and evolution of the global p-mode oscillations at the photosphere.}
The Alfv\'en waves/pulses are also put forward as a trigger of the spicular jets in the chromosphere \cite{1982ApJ...257..345H,1999ApJ...514..493K,2022MNRAS.511.4134S}. Hollweg \cite{1982ApJ...257..345H} 
proposed  a MHD model for the formation and evolution of spicule-like cool jets using an Alfv\'en pulse of the magnitude of 1 km s$^{-1}$ at the photosphere. 
Recently, Singh B et al. \cite{2022MNRAS.511.4134S}  delineated the comprehensive physical scenario about the origin and motions of the cool spicular jets in the ideal, single fluid MHD  by implementing large-amplitude Alfv\'en pulse at the chromospheric height. These  jets show quasi-periodic rise and fall in the model solar atmosphere and describe the physical manifestation of spicules. There are several single fluid MHD model in ideal and non-ideal regime, that clearly depicts the generation of cool spicular jets \cite{2017A&A...597A.133K,2021ApJ...913...19M}.

Additionally, two fluid numerical models understand the multi-fluid physical effects and their roles as compared to single fluid models. Therefore, two fluid models are one of the more convincing methods providing the new insight of the triggering and physical properties of such spicule-like jets. The partially ionized solar chromosphere contains ions, electrons and neutrals. Many models of the solar chromosphere, therefore, include non-ideal factors like conduction, radiative cooling, and dissipation predominantly caused by the ion-neutral collisions, providing crucial information regarding the heating and plasma dynamics in the chromosphere\cite{2018SSRv..214...58B,2020A&A...635A..28W,2021MNRAS.506..989K}. 
James et al. \cite{2003A&A...406..715J} have described the generation of spicules caused by the dissipation of high-frequency Alfv\'en waves/pulse invoking the physics of ion-neutral collision damping in axisymmetric and initially untwisted flux tubes by performing 1.5-D simulation. 
The dissipation of waves attributed by ion-neutral collisions and thermal conduction, in general take place to deposit the additional energy in the solar chromosphere that subsequently takes into account the evolution of spicular cool jets \cite{2012ApJ...747...87K,2013ApJS..209...16S,2017PPCF...59a4038K,2017ApJ...849...78K,2019A&A...630A..79P}.

In this work, we have carried out a 2.5-D numerical simulation in two-fluid regime. In the presence of ions and neutrals along with ion-neutral collisions and thermal conduction we provide a perspective of the triggering of cool spicule-like jets triggered initially by the implementation of Alfv\'en pulse at the solar chromosphere. We impose a non-linear, impulsive Alfv\'en pulse in the solar chromosphere to drive the cool spicular jets in the realistic temperature { and} gravitationally stratified model of the vetically structured solar atmosphere.  
Previously, \cite{2022MNRAS.511.4134S} proposed a model of the spicule-like cool jets in ideal, single-fluid MHD by implementation of randomly generated multiple Alfv\'en pulses of magnitudes 50-90 km s$^{-1}$ in the solar chromosphere . In the present work, we revisit this previously proposed case of ideal MHD, and here we perform a 2.5-D numerical model for the evolution and triggering of cool jets in two-fluid (ions \& neutrals) regime of the chromospheric plasma taking into account thermal conduction and ion-neutral collision 
using JOANNA code. We find that the existence of ion-neutral collisions and thermal conduction influence the evolution and kinematics of such spicular jets triggered initially by non-linear Alfv\'en pulses in the chromosphere. These physical candidates are required for the formation of such spicular cool jets in the solar chromosphere.
In this paper, we investigate the triggering mechanism and evolution of these cool jets which closely match with the observations. This paper is put in the order as follows. The basic two-fluid model and initial setup of the solar atmosphere are presented in Section 2. The numerical simulation and its results are described in Section 3. The discussion and conclusions of this work are presented in Section 4.

\section{Model and initial configuration}

\subsection{The Two-Fluid Model}
 { In this model, we consider} the two-fluid system consisting of ions and neutrals along with ions-neutral collisions. We do not consider the consequences of recombination and ionisation of ions and neutrals. We also take into account the thermal conduction which contributes in the
formation/dynamics of the cool jets. The governing equations of two-fluid system in the solar magnetized plasma { are given} as follows
\cite{2011A&A...534A..93Z,2012ASPC..454..129S,2012ApJ...747...87K,2012PhPl...19g2508M,2017ApJ...849...78K,2014SSRv..184..107L,2016ApJ...818..128O,2017ApJ...836..197M,2019A&A...627A..25P,2021MNRAS.506..989K,2022Ap&SS.367..111M}:

\begin{equation}
\frac{\partial \varrho_{\rm i}}
{\partial t}+\nabla\cdot(\varrho_{\rm i} \mathbf{V}_{\rm i}) = 0, 
\label{eq:ion_continuity} 
\end{equation}

\begin{equation}
\frac{\partial \varrho_{\rm n}}{\partial t}+
\nabla\cdot(\varrho_{\rm n} \mathbf{V}_{\rm n}) = 0 , 
\label{eq:neutral_continuity} 
\end{equation}


\begin{equation}
\label{eq:ion_momentum}
\begin{split}
\begin{aligned}
&\varrho_{i}\left(\frac{\partial{{\bf V} _{i}}}{\partial t} + ({{\bf V} _{i}}\cdot \nabla){{\bf V} _{i}}\right)=-\nabla p_{i}+\frac{1}{\mu} (\nabla \times {\bf B}) \times {\bf B}+\varrho_{i}{\bf g}  \\
&-\alpha_{c}({{\bf V}_{i}}-{{\bf V}_{n}}) \, , 
\end{aligned}
\end{split}
\end{equation}

\begin{equation}
\label{eq:neutral_momentum}
\begin{split}
\begin{aligned}
&\varrho_{n}\left(\frac{\partial{{\bf V} _{n}}}{\partial t} + ({{\bf V} _{n}}\cdot \nabla){{\bf V} _{i}}\right)=-\nabla p_{n} +\varrho_{n} {\bf g} \\
&+\alpha_{c}({{\bf V}_{i}}-{{\bf V}_{n}}) \, , 
\end{aligned}
\end{split}
\end{equation}

\begin{equation}
\frac{\partial \mathbf{B}}{\partial t}
=
\nabla \times (\mathbf{V_{i} \times } \mathbf{B})
\, ,
\hspace{6mm}
 \nabla \cdot{\mathbf B} = 0
 \, .
\label{eq:ions_induction} 
\end{equation}


\begin{equation}
\begin{split}
& \frac{\partial E{\rm {\rm _i}}}{\partial t}+\nabla\cdot \left[\left(E{\rm {\rm _i}}+p{\rm {\rm _{i}}}+\frac{{\mathbf{B}}^2}{2\mu}\right)\mathbf{V_{\rm i}}-\frac{\mathbf{B}}{\mu}(\mathbf{V_{\rm i}} \cdot {\mathbf{B}})\right]  = \\
 &  
  Q{\rm _i} + (\varrho{\rm {\rm _i}} \mathbf{g} + \alpha_{\rm c} (\mathbf{V_{\rm n}}-\mathbf{V_{\rm i}})) \cdot \mathbf{V_{\rm i}} + \nabla \cdot \mathbf{q_i} \, ,\\
\end{split}
\end{equation}

\begin{equation}
\begin{split}
 & \frac{\partial E_{\rm n}}{\partial t}+\nabla\cdot((E_{\rm n}+p_{\rm n})\mathbf{V_{\rm n}}) =  Q_{\rm n} +
 (\varrho_{\rm n} \mathbf{g}  - \alpha_{\rm c} (\mathbf{V_{\rm n}}-\mathbf{V_{\rm i}} )) \cdot \mathbf{V_{\rm n}} + \nabla \cdot \mathbf{q_n}  \, , \\
 \end{split}
\end{equation}

Here
\begin{align}
E_{\rm i} = \frac{p_{\rm i}}{\gamma-1} + \frac{1}{2}\varrho{\rm {\rm _i}} |\mathbf{V{\rm _i}}|^2 + \frac{ |\mathbf{B}|^2}{2\mu},
\end{align}
and

\begin{align}
E_{\rm n} = \frac{p_{\rm n}}{\gamma-1} + \frac{1}{2}\varrho_n |\mathbf{V_{\rm n}}|^2\, ,
\end{align}






Here, the { subscripts} 'i' and 'n' depict respectively the components of ions and neutrals. The symbols ${\varrho_{\rm i}}_{,\rm n}$ represent the mass densities, ${p_{\rm i}}_{,\rm n}$ { thermal} pressures, and ${{V}_{\rm i}}_{,\rm n}$ is the plasma velocity. { B} denotes the magnetic field. The constant $\mu$ appear for the magnetic permeability. We consider the specific heats ratio represented by $\gamma$, which is equal to 5/3. 
The g=[0,-g,0] depicts the acceleration due to gravity at the Sun. We fix the value of 'g' equal to 274.78 m s$^{-2}$ that is an average value at { the} solar surface i.e., photosphere. The Hall term electron pressure has been neglected in ideal magnetic induction equation (Eq.~2.5), and the diffusion is numerical. The E$_{i}$ and E$_{n}$ expressed as total  energy densities.
The coefficient $\alpha_c$ 
denotes the ion-neutral collisions expressed as follows \cite{1965RvPP....1..205B,2011A&A...529A..82Z,2021MNRAS.506..989K}: 

\begin{equation}
\alpha_{\rm c} = \frac{4}{3} \frac{\sigma_{\rm in}\,\varrho_{\rm i}\,\varrho_{\rm n}}{m_H(\mu_{\rm i}+\mu_{\rm n})} \sqrt{\frac{8k_B}{\pi m_H} \left(\frac{T{\rm _i}}{\mu_{\rm i}}+\frac{T_{\rm n}}{\mu_{\rm n}}\right)} \, .
\end{equation}

Here, $m_{H}$ represents the hydrogen mass, $k_{B}$ is the Boltzmann's constant, the symbols $\mu_{i,n}$ represent the mean mass of the ions and neutrals, respectively. We fix the value of $\mu_{i}$ = 0.58 for ions+electrons and $\mu_{n}$ = 1.21 for neutrals from the given solar abundance  \cite{2021MNRAS.506..989K}. The symbols $T_{i,n}$ denote the temperature of ions and neutrals, respectively. The symbol $\sigma_{in}$ represents { the} cross-section of ion-neutral collision as a quantum value equal to 1.4 $\times$ $10^{-15}$ m$^{2}$ for momentum exchange among the particles \cite{2013A&A...554A..22V, 2021MNRAS.506..989K}.

{ The additional terms in the energy equations 'Q${_i}$' and 'Q$_{n}$' represent the frictional interaction between particles resulting generates the additional heat, which are expressed as:}

\begin{equation}
Q{\rm _i} = \alpha_{c} \left[ \frac{1}{2} \vert\mathbf{V{\rm _i}}-\mathbf{V_{\rm n}}\vert ^2 - \frac{3}{2} \frac{k_B}{m_{\rm H}(\mu_{\rm i}+\mu_{\rm n})} (T{\rm _i}-T_{\rm n}) \right]\, , \\
\end{equation}
\begin{equation}
Q_{\rm n} = \alpha_{c} \left[ \frac{1}{2} \vert\mathbf{V{\rm _i}}-\mathbf{V_{\rm n}}\vert ^2 - \frac{3}{2} \frac{k_B}{m_{\rm H}(\mu_{\rm i}+\mu_{\rm n})} (T_{\rm n}-T{\rm _i}) \right]\, ,\\
\end{equation}




 { In this model, the divergence of the embedded magnetic field lines ($B$) was kept small.} Here, we do not take into account some non-ideal dissipative effects such as viscosity, resistivity, plasma cooling and/or heating, magnetic diffusivity etc.
However, we implement in the model the effect of thermal conduction by figuring out { the} heat flux vector (q$_{i,n}$). 
In case of neutrals, thermal conduction is isotropic which is expressed { as follows}: 

\begin{equation}
{\mathbf q}_{\rm n} = \kappa_{\rm n} \nabla T_{\rm n}\, . 
\end{equation}

Here, $\kappa_{\rm n}$ depicts the coefficient of thermal conduction in case of neutrals which is expressed as \cite{2007ApJS..171..520C,2022Ap&SS.367..111M} 

\begin{equation}
\kappa_{\rm n} = 
\frac{29.6\, T_{\rm n}}{1+\sqrt{7.6\cdot 10^5\, {\rm K} /T_{\rm n}}} 
\frac{\mu_{\rm n} m_{H}}{k_{\rm B}}\, , 
\end{equation}

where $k_{\rm B}$ is $1.3807\times 10^{-16}\, {\rm cm}^2\, 
{\rm g}\, {\rm s}^{-2}\, {\rm K}^{-1}$. 

Thermal conduction for the ions, is highly anisotropic that permits the flow of heat through the length of magnetic field lines. However, the heat flux vector across the length of the magnetic field lines is negligibly small. The expression for the heat flux ($q_{i}$) is expressed as

\begin{equation}
{\mathbf q}_{\rm i} = \kappa_{\parallel}{\mathbf b} \nabla({\mathbf b}\cdot T_{\rm i}) \, , 
\end{equation}

where ${ b}={ B}/B$ depicts the unit vector { aligning} in the direction of the magnetic field. The coefficient of thermal conduction $\kappa_{\parallel}$ is expressed as \cite{1962pfig.book.....S, 2022Ap&SS.367..111M}:

\begin{equation}
\kappa_{\parallel} \approx 4.6\cdot 10^{13} 
\left(
\frac{T_{\rm e}}{10^8\, {\rm K}}\right)^{5/2}
\frac{40}{\Lambda_{c}}\, 
\hspace{3mm}
[{\rm erg}\, {\rm s}^{-1}\, {\rm cm}^{-1}\, {\rm K}^{-1}]  
\end{equation}

Here, $\Lambda_{c}$ denotes the quantum Coulomb logarithm which is expressed as \cite{2013JaJAP..52j8002H, 2022Ap&SS.367..111M} 

\begin{equation}
\Lambda_{c} \approx 30.9 - \log\frac{n_{\rm e}^{1/2}}{T_{\rm e}k_{\rm B}^*}\, .
\end{equation}
Here, the Boltzmann constant $k_{\rm B}^*$ is expressed in ${\rm eV}\,$ K$^{-1}$. The symbol $n_{e}$ represents the electron particle { number} density, while $T_{e}$ is the electron temperature as expressed in eV.

\begin{figure*}
\centering
\includegraphics[width=6.5cm, height=8.5cm]{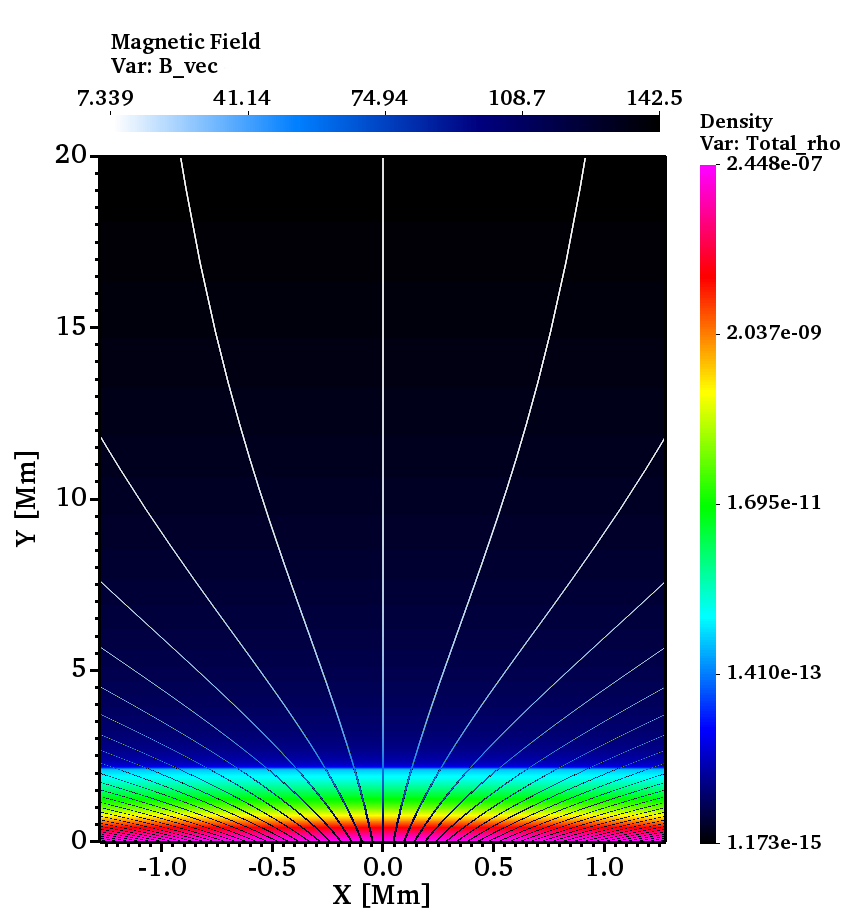}
\includegraphics[width=6.5cm, height=5.5cm]{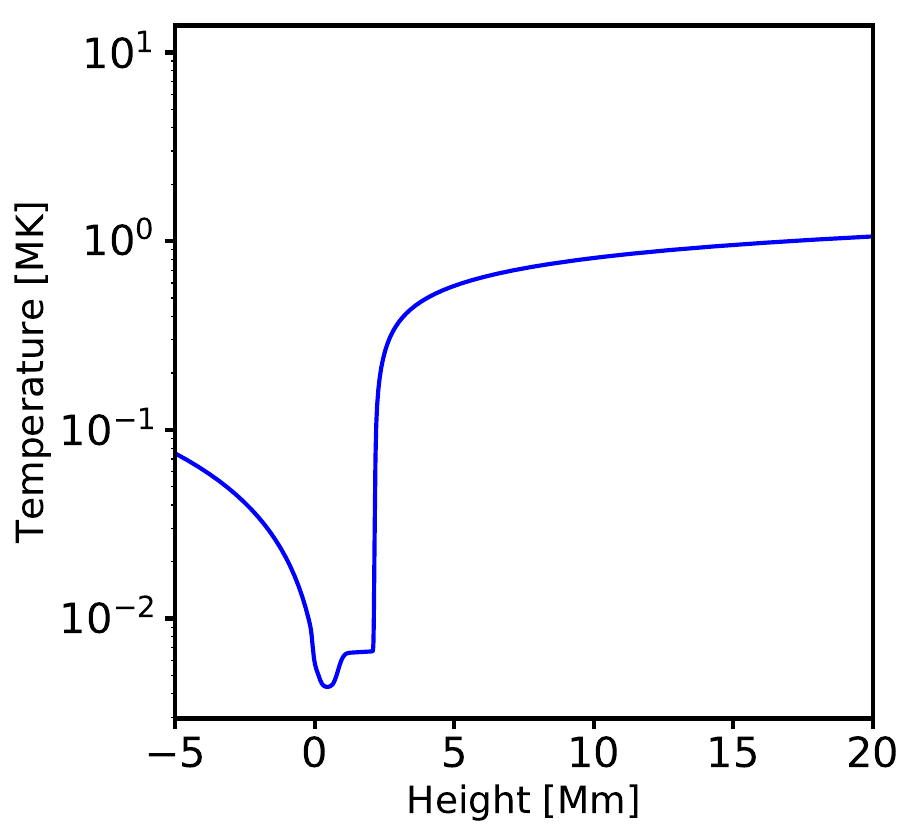}
\caption{ The left image displays smoothly curved and expanding magnetic field lines permeating the gravitionally stratified solar atmosphere maintained initially at an equilibrium from the solar photosphere upto the inner corona. { The density and magnetic field strength in colorbar are given in g cm$^{-3}$ and Gauss respectively}. The right panel shows the temperature variation as a function of height in this vertically (or longitudinally) structured  localized solar atmosphere as theorized by Avrett \& Loeser \cite{2008ApJS..175..229A}.}
\label{Magnetic_field_Temp}
\end{figure*}



\begin{figure*}
\centering
\mbox{
\includegraphics[width=4.5cm,height=3.5cm]{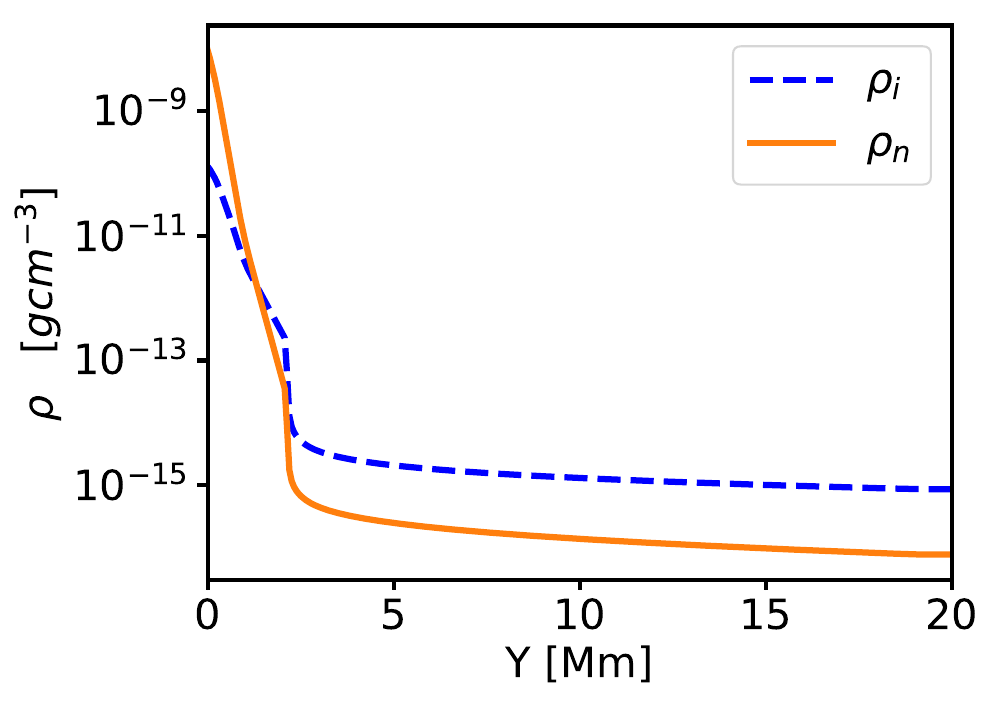}
\includegraphics[width=4.5cm,height=3.5cm]{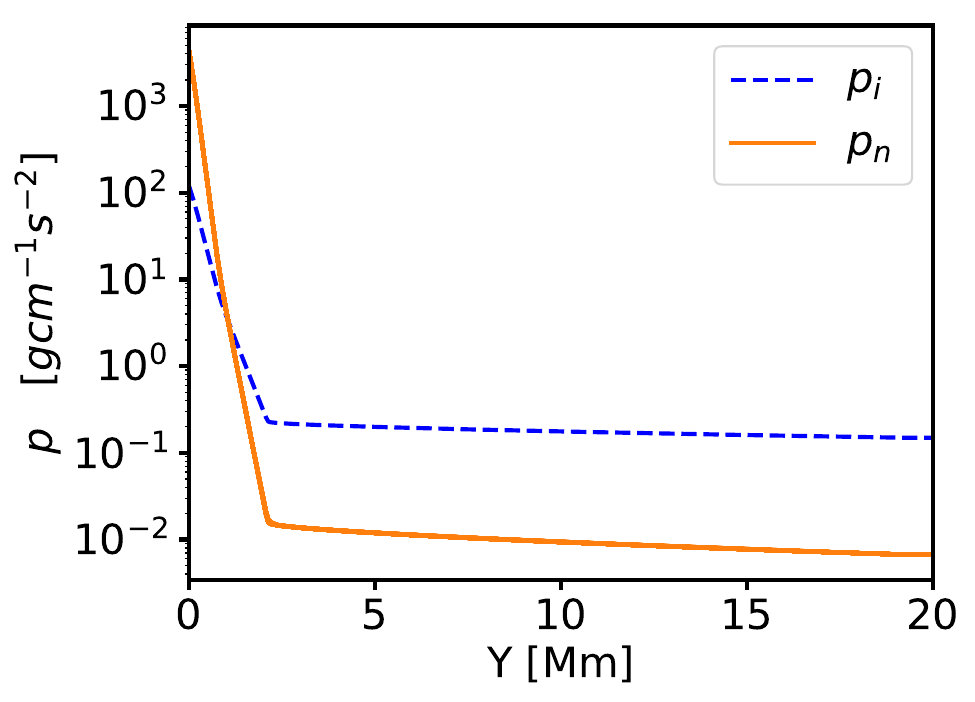}
\includegraphics[width=4.5cm,height=3.5cm]{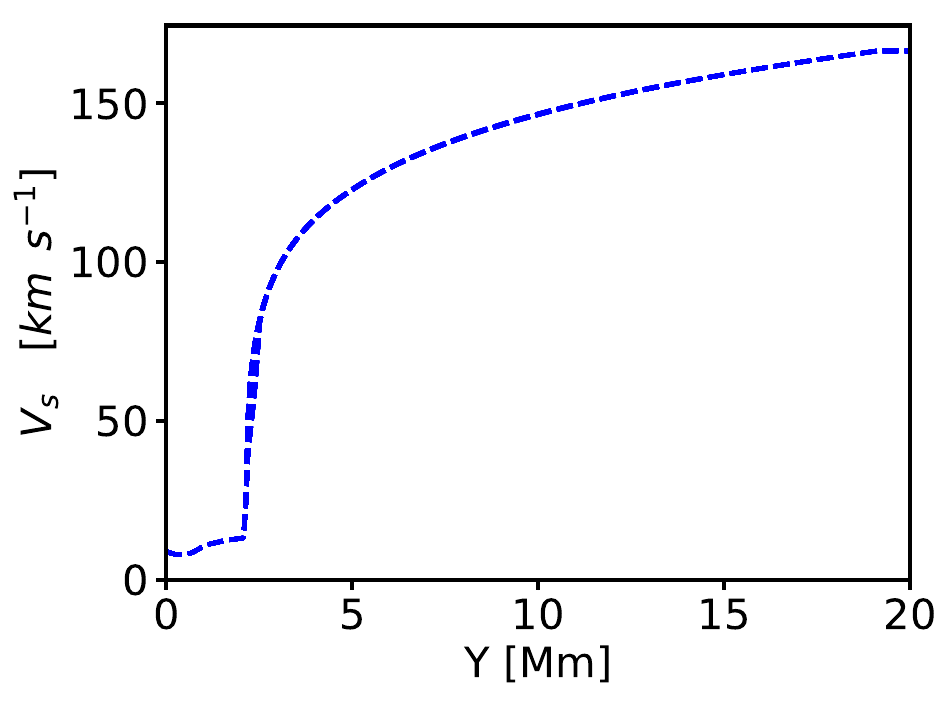}
}

\mbox{
\includegraphics[width=4.5cm,height=3.5cm]{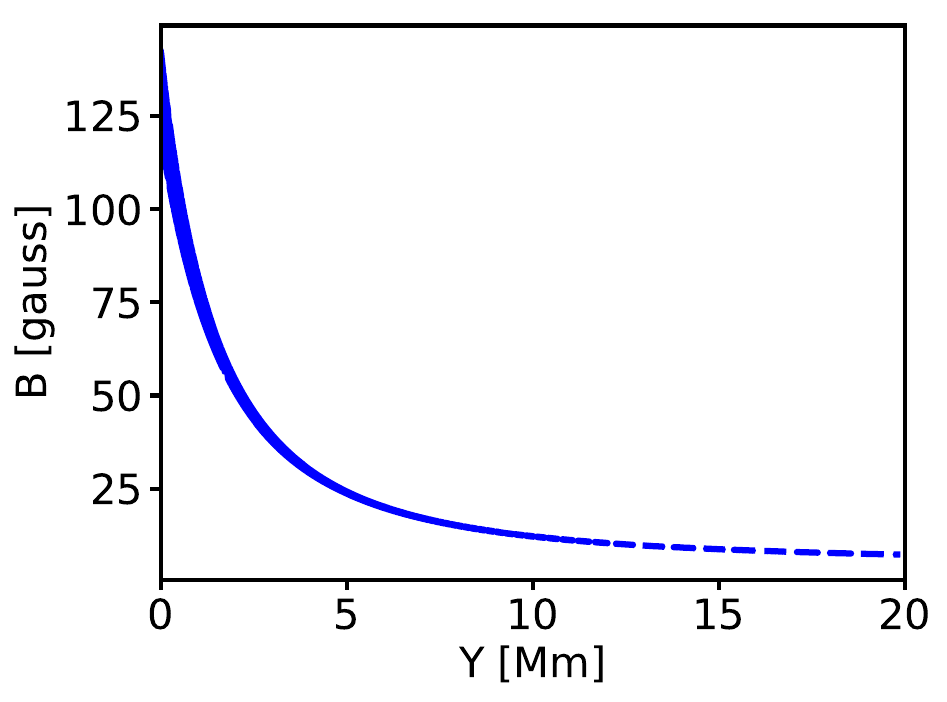}
\includegraphics[width=4.5cm,height=3.5cm]{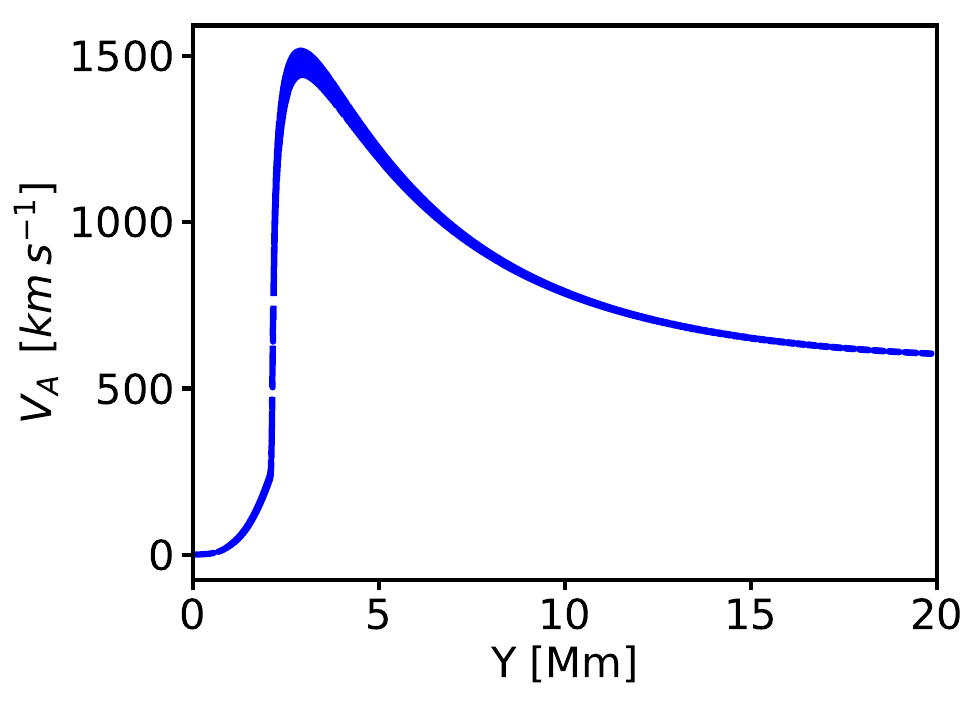}
\includegraphics[width=4.5cm,height=3.5cm]{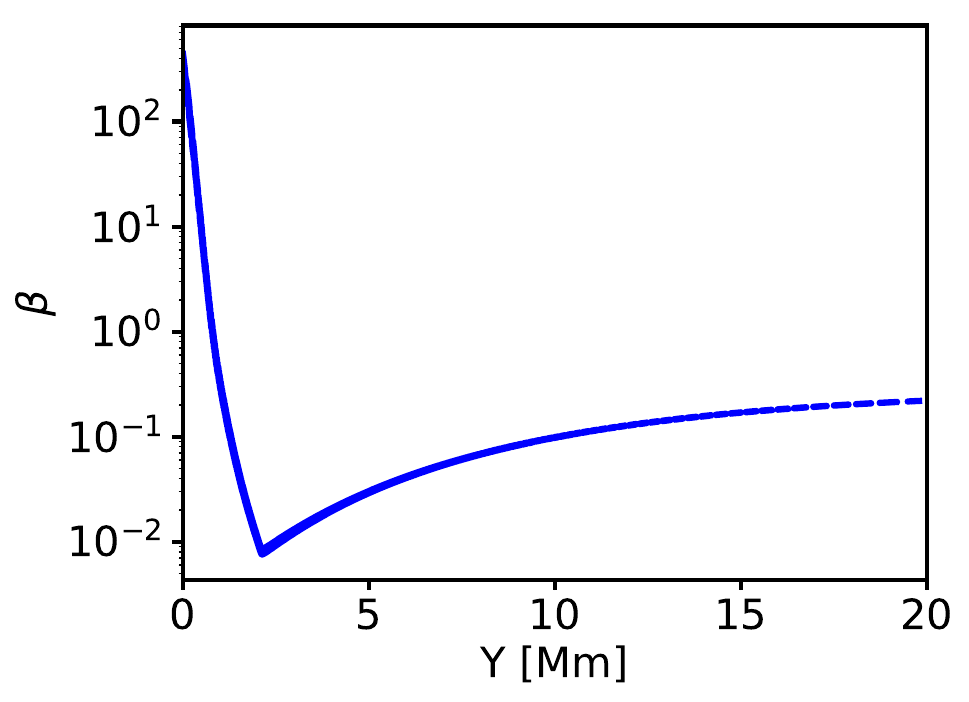}
}

\caption{The physical background parameters vs height. Top left panel shows respectively the mass density for ions (blue-curve) and neutrals (magenta-curve). The top-middle panel demonstrates respectively the plasma pressure for ions (blue-curve) and neutrals (magenta-curve). The sound speed (top-right), magnetic field (bottom-left), Alfv\'en speed (bottom-middle), and plasma beta (bottom-right) { are also displayed}.}
\label{}
\end{figure*}


\subsection{Equilibrium Configurations of the Model Solar Atmosphere}

Initially, the considered model atmosphere extending from { the} solar photosphere to the inner corona is in equilibrium state. { As the average flow at the transition region is estimated to be of the order of about 3 km s$^{-1}$ (e.g. Avrett \& Loeser \cite{2008ApJS..175..229A}), initially (at $t$=0 s) we set the quasi-static ($V_{i}$=$V_{n}$=0) atmosphere. } 
In addition, the Lorentz force affects only the ions while neutrals are obviously remain unaffected. Therefore, the equations \ref{eq:ion_momentum} and \ref{eq:neutral_momentum} can be written as follows: 

\begin{equation}
0 = - \nabla p_{i} + \rho_{i}\boldsymbol{g} + \frac{1}{\mu}(\nabla\times\boldsymbol{B})\times\boldsymbol{B} 
\end{equation}
and
\begin{equation}
0 = - \nabla p_{n} + \rho_{n}\boldsymbol{g}  
\end{equation}

We take into account an appropriate magnetised solar atmosphere that constitutes the region between convection zone (-3.0 Mm) and corona (20 Mm). The model solar atmosphere follows the hydrostatic equilibrium for ions initially immersed in the force free and current free magnetic field given as
\begin{equation}
(\nabla \times \textbf{B}) \times \textbf{B} = 0, \hspace{2cm} \nabla \times \textbf{B} = 0.
\end{equation}

The horizontal ($\boldsymbol{B}_x$) { component}, vertical ($\boldsymbol{B}_y$) { component}, and transversal ($\boldsymbol{B}_z$) { component} of the equilibrium magnetic field are expressed as per { the B.C. Low model}: \cite{1985ApJ...293...31L, 2022MNRAS.511.4134S}
\begin{equation}
    B_x(x,y)=\frac{-2S(x-a)(y-b)}{((x-a)^2+(y-b)^2)^2},
\end{equation}
\begin{equation}
        B_y(x,y)=\frac{2S(x-a)^2-S((x-a)^2+(y-b)^2)}{((x-a)^2+(y-b)^2)^2} + B_v,
\end{equation}
\begin{equation}
     B_z(x,y)= 0.
\end{equation}

Here, the symbol 'S' { denotes} the magnitude of the magnetic pole. The variables 'a' and 'b' respectively represent the horizontal and vertical location of this singular magnetic pole. We choose its magnitude as S $\simeq$1240 Gauss in the sub-photospheric layer, i.e., convection zone, at the position of (a,b)=(0,-3.0) Mm. We also fix the additional magnetic field B$_{v}$=5 Gauss in the solar corona (i.e., y$_{ref}$=50 Mm) which forms quiet-Sun atmosphere with sufficiently strong magnetic field. The magnetic line of forces come out from the chosen pole in the convection zone and fan-out through the photosphere in the inner corona as displayed in the left-panel of Fig.~1. The strength (or magnitude) of the magnetic field falls-off with height in corona according to  Eqs. 2.17, 2.18 and 2.19.

The modelled solar atmosphere is also maintained by the force-free condition in the background. Therefore, the gravity force equalises the pressure gradient force, making the medium approximately force-free. It is possible to express it as follows \cite{Priest_2014, 2021MNRAS.506..989K}:

\begin{equation}
{p_{\rm i}}_{,\rm n}= p_{0 i,n} \exp{\left(-\int_{y_{ref}}^{y}{\frac{dy'}{\Lambda(y')}}\right )},
\end{equation}

\begin{equation}
    {\varrho_{i,n}}= \frac{{p_{\rm i}}_{,\rm n}}{g\Lambda(y)}, 
\end{equation}
\\
where,
\begin{equation}
    \Lambda(y)=\frac{k_B  T_{i,n}}{mg},
\end{equation}
\begin{equation}
    p_{i,n} = \frac{\rho_{i,n} T_{i,n} k_{B}}{m_{H} \mu_{i,n}}.
\end{equation}


where, $ p_{0 i,n}$ representing the thermal pressure of ions as well as neutrals at the given reference point (i.e., $y_{ref}$ = 50 Mm). We fix the value of $p_{0i}$ = 10$^\mathrm{-1}$ dyn cm$^\mathrm{-2}$ and $p_{0n}$ = 3 $\times$ 10$^\mathrm{-3}$ dyn cm$^\mathrm{-2}$ at the $y_{ref}$=50 Mm in the inner corona. { Initially (at t=0), we assume that $T_{i}(y)$=$T_{n}(y)$ but later on they are allowed to evolve according to the two-fluid model.}

Initially, we adopt the realistic temperature profile which is theorized by Avrett \& Loeser \cite{2008ApJS..175..229A}, using various line profiles recorded at multiple heights (regions) of the solar atmosphere (see right panel of Fig.~1). 
The mosaic of Fig.~2 shows the  density (top left), thermal pressure (top middle), sound speed (top right), magnetic field (bottom left), Alfv\'en speed (bottom middle), and plasma beta (bottom right) as a function of height in the gravitationally stratified and magnetized solar atmosphere. It depicts the longitudinal (or vertical) structuring of { the model soalar atmosphere}. The mass density { of the plasma gas} and { its} thermal pressure for ions ($p_{i}$) and neutrals ($p_{n}$) at different heights, are estimated using equations Eqs.~2.20-2.23. 
The detailed descriptions as given above demonstrate that we employ a realistic, magnetically and gravitationally stratified model of the solar atmosphere to understand the generation of cool spicule-like jets.

\subsection{Numerical Methods}

We numerically solve the complete set of two-fluid equations within JOANNA code \cite{2017ApJ...849...78K, 2021MNRAS.506..989K} by carrying out the { implementation of} initial and boundary conditions. To assure stability and accuracy in the present simulation, we fix the Courant–Friedrichs–Lewy (CFL) number as 0.3 and use 2nd-order Runge-Kutta (RK2) numerical technique for time stepping because it is a self-starting method for solving the differential equations. The source terms such as ion-neutral drag force and thermal conduction are stiff and therefore they require an explicit treatment 
\cite{2017ApJ...849...78K, 2021MNRAS.506..989K}. Furthermore, we have employed Harten-Lax-van Leer discontinuities (HLLD) as a Riemann solution to improve shock stability and low-speed flow accuracy. We set the model solar atmosphere of $2.56$ Mm ($-1.28\leq x\leq1.28$) Mm in $x$, and 20 Mm in $y$ ($0\leq y\leq20$) Mm. We hold the static uniform grid from 0 Mm up to 5.12 Mm in the $y$-direction, and fix the stretched grid above y=5.12 Mm up to y=20 Mm. The spatial resolution in the zone of uniform grid is 10 km. 
In the upper part of the model solar atmosphere, the proper stretched grid has been employed which acts like a sponge, effectively soaking up the incoming signals and minimizing the amount of reflection from the top boundary. As a result, there is no substantial reflected signals { are seen} in the system, particularly in the top zone of the numerical domain. The majority of the energy exchange from Alfv\'en pulse occurs in the chromosphere-TR-inner corona below y=5.0 Mm, and cool jets propagate in that particular region { only}. 
We impose all four boundaries on the simulation box to ensure that all the boundaries continue to preserve the equilibrium in { the} physical parameters. { More specifically, at the bottom and top boundaries of the numerical domain, we hold fixed all plasma quantities to their equilibrium values and at the side boundaries we use open boundary conditions. We found that top and bottom boundaries minimize numerically induced reflections of the incoming signal, and the open side boundaries allow the signal to leave the simulation box.} 
{ It} is the appropriate way to have boundaries that function relatively well when gravity is present in the system.
We run the numerical simulations for 1000 s utilizing multiple passage interface (MPI) and save the output files every 5 s.

\subsection{Perturbations}

At an initial time of the present two-fluid simulation i.e., at $t$= 0 s, we impose the transverse velocity pulse $V_{z}$ in ions and neutrals both, which mimic the Alfv\'en pulse as given by the following eq. \cite{2022MNRAS.511.4134S}:

\begin{equation}
V_{z} = A_{v} \times exp \left(-\frac{(x - x_{0})^2 + (y - y_{0})^2}{w^2}\right) .
\end{equation} 

In above Eq., $A_v$ appears for the amplitude of the applied driver. The symbols $x_0$ and $y_0$ are respectively the horizontal and vertical positions of the pulse, while 'w' is its width, { which is taken as 50 km}. We impose the amplitude of Alfv\'en pulse ($A_v$)= 150 km s$^\mathrm{-1}$ at $(x_0,y_0)$ = (0, 1.5) Mm, which corresponds to non-linear { and impulsive} in nature. 
The applied amplitude of the Alfv\'en pulse is strong enough to trigger cool jets { via energy exchange through ponderomotive force}. It is chosen in the light of { previously} existing knowledge that non-linear, large amplitude transversal perturbations  and accompanying Alfv\'en wave propagation have already been observed in the localized solar atmosphere \cite{ 2007Sci...318.1574D}. In 2.5-D regime, we implement the transverse Alfv\'en pulse by applying velocity pulse in the ignorable horizontal z-direction.
The spicule-like structures in the chromosphere were observed to carry the large amplitude Alfv\'en motions on their spine  \cite{2007Sci...318.1574D}. Such impulsive motions can give a sling-shot transverse pulse (an Alfv\'enic whip-like motion) or periodic wave-like motions to the flux tubes, leading to similar physical conditions as those modeled in the present paper.~ 
However, on the contrary, some other observations do not reveal such high, non-linear, and impulsive velocity perturbations in waves, e.g., by measuring line widths, for instance \cite{2011ApJ...736L..24O,2017Sci...356.1269M}. The present model has extreme amplitudes larger than the above mentioned observations and future work should address this. So, we conclude that the driver that we choose in the present work is an episodic driver which can launch spicules or other cool jets through impulsive processes occurring in the localized solar chromosphere. In the non-ideal and two fluid plasma where ion neutral collision and thermal conduction were at the work, the high amplitude of the pulse in our case has only generated the ponderomotive force in the chromosphere, and triggered the field aligned perturbations. Therefore, the presented physical model should be considered as an impulsive and episodic origin of cool jets where reconnection generated velocity perturbations are likely formed in the localized lower solar atmosphere.


\begin{figure*}
\centering
\mbox{
\includegraphics[width=2.7cm, height=4.5cm]{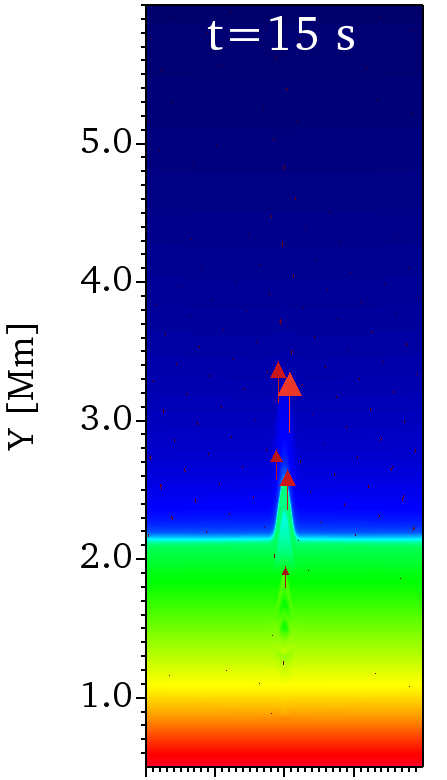}
\includegraphics[width=1.8cm, height=4.5cm]{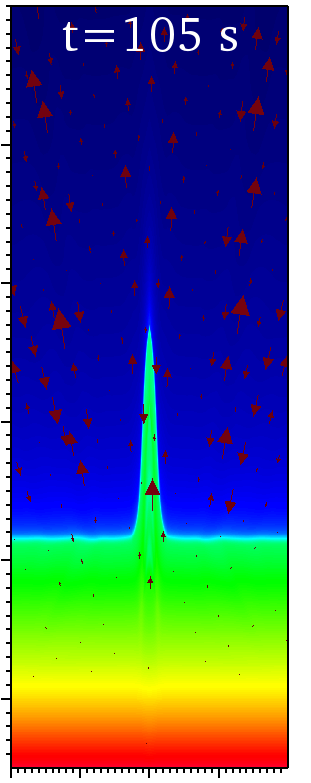}
\includegraphics[width=1.8cm, height=4.5cm]{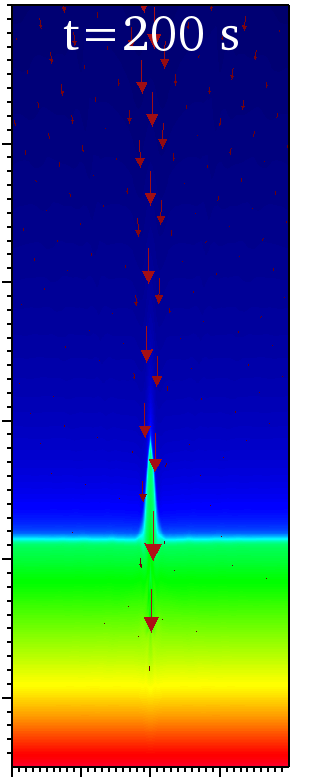}
\includegraphics[width=1.8cm, height=4.5cm]{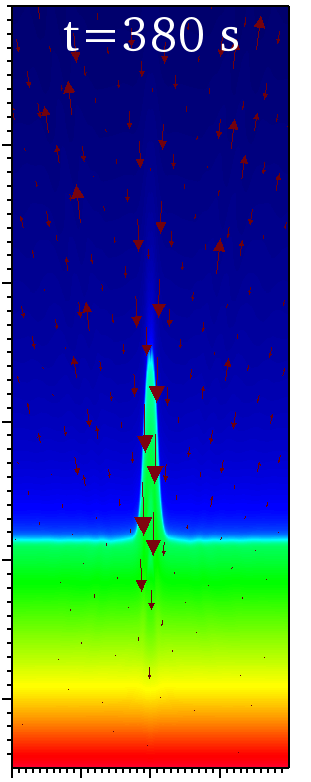}
\includegraphics[width=1.8cm, height=4.5cm]{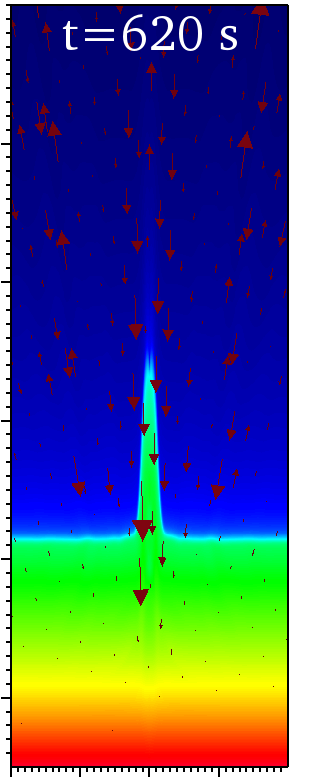}
\includegraphics[width=3.0cm, height=4.5cm]{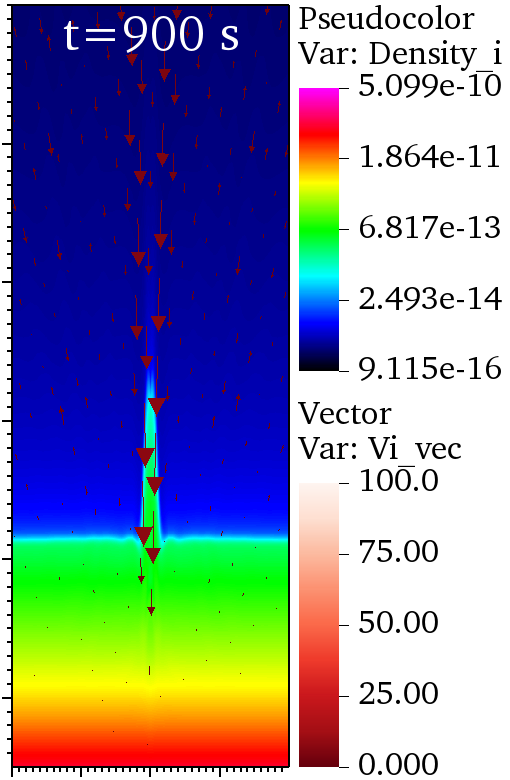}
}
\mbox{
\includegraphics[width=2.6cm, height=5cm]{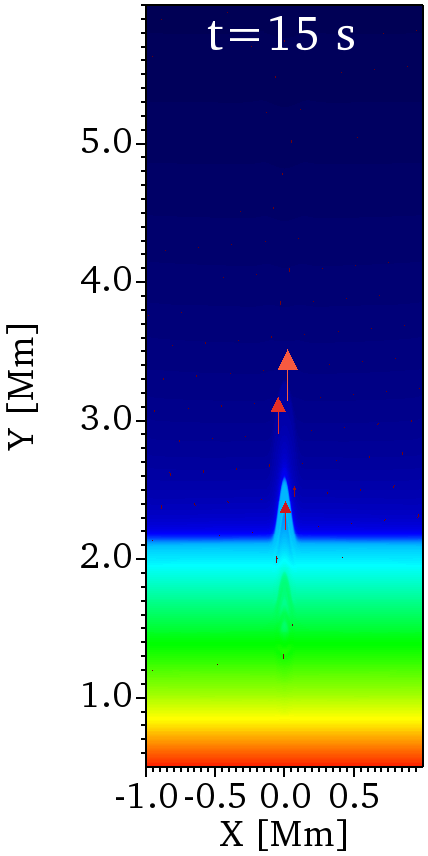}
\includegraphics[width=1.8cm, height=5cm]{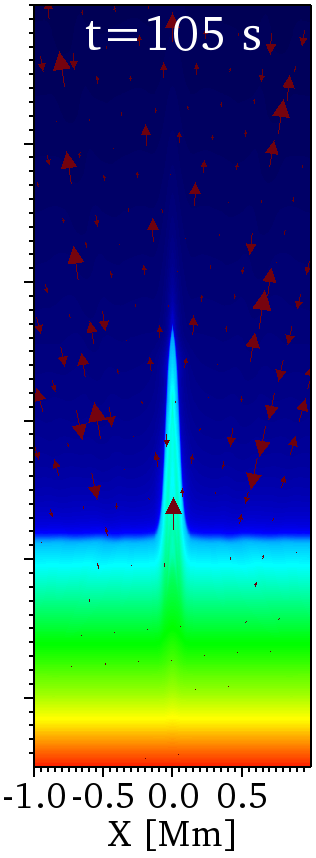}
\includegraphics[width=1.8cm, height=5cm]{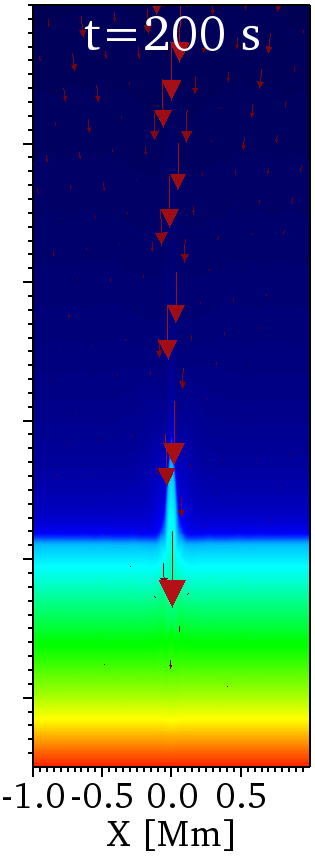}
\includegraphics[width=1.8cm, height=5cm]{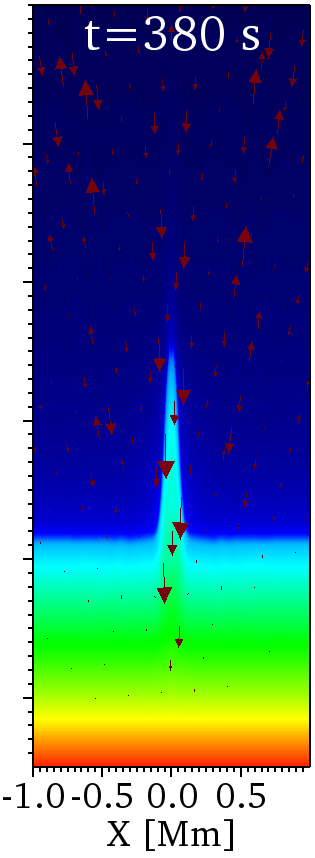}
\includegraphics[width=1.8cm, height=5cm]{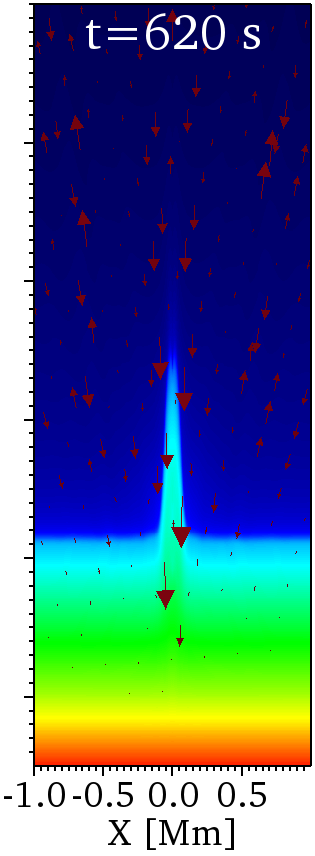}
\includegraphics[width=3.2cm, height=5cm]{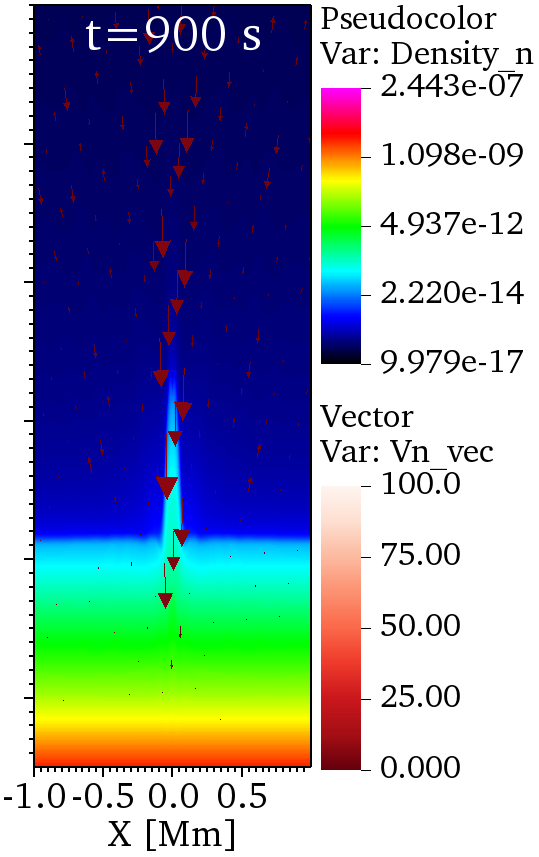}
}

 \caption{Spatial-temporal variation of density maps { in g cm$^{-3}$,} for ions (upper panels) and for neutrals (bottom panels) during launch of cool jets. These jets are triggered by implementing the non-linear transverse pulse ($V_z$)  mimicking Alfv\'en pules and  implemented in the chromosphere at y=1.5 Mm. This pulse produces initially field-aligned perturbations followed by { recurrent} magnetoacoustic shocks. These shocks create collimated mass transports as cool jets. The red arrows depict the velocity vectors in km s$^{-1}$, which are overlaid on each density map for ions (upper panels) and neutrals (bottom panels). These velocity vectors shows the upward and downward propagation of jet's plasma in the model solar atmosphere.}
    \label{}
\end{figure*}

 
\begin{figure*}
\centering
\mbox{
\includegraphics[width=12cm,height=8.0cm]{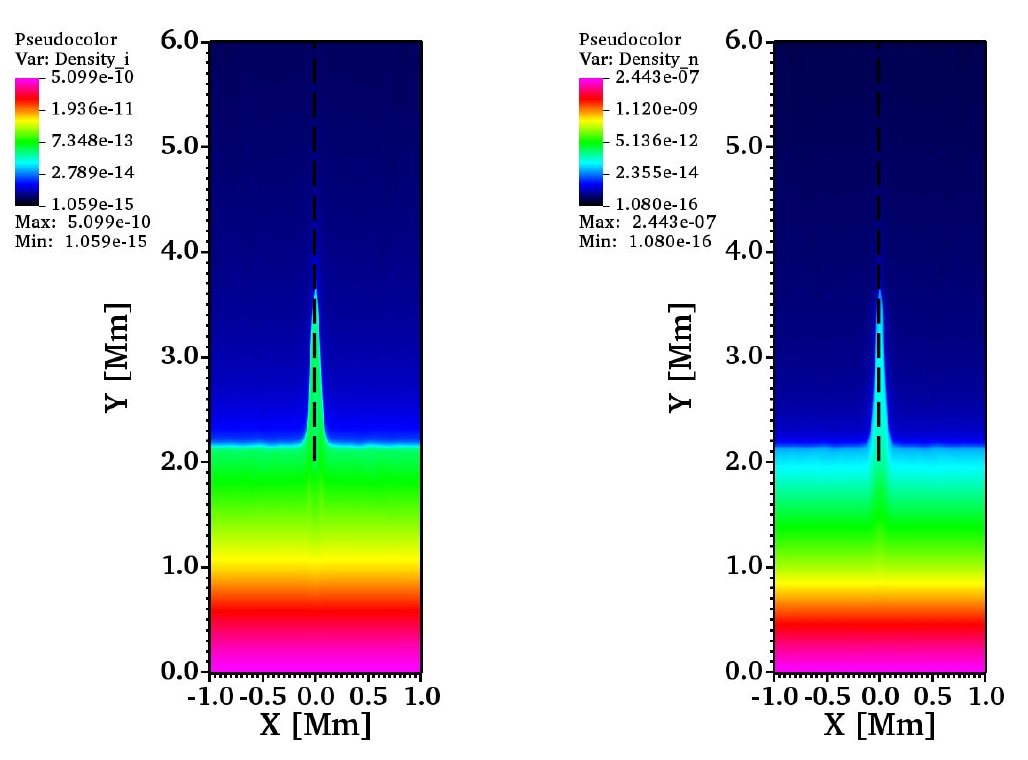}
}
\mbox{
\includegraphics[width=7.0cm,height=7cm]{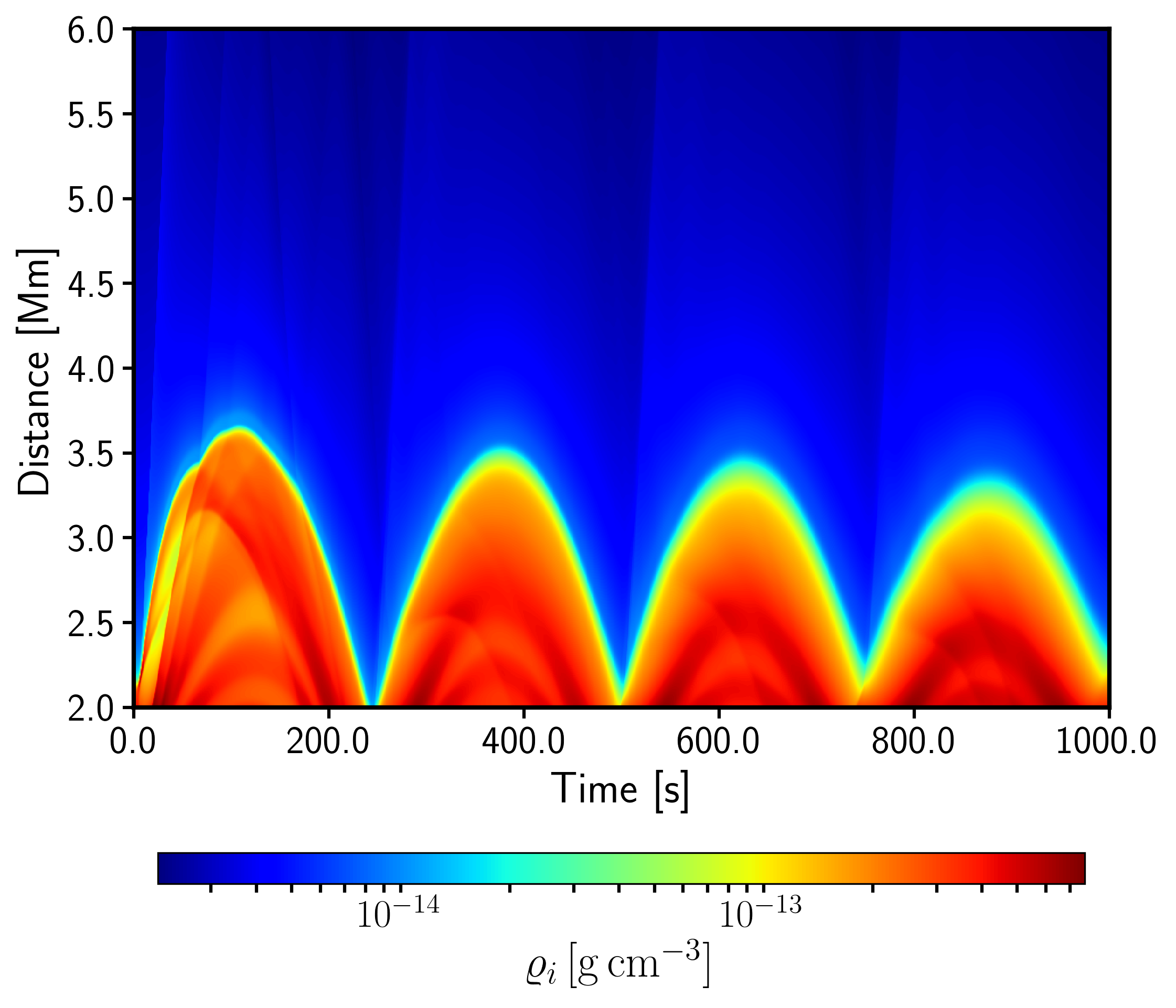}
\includegraphics[width=7.0cm,height=7cm]{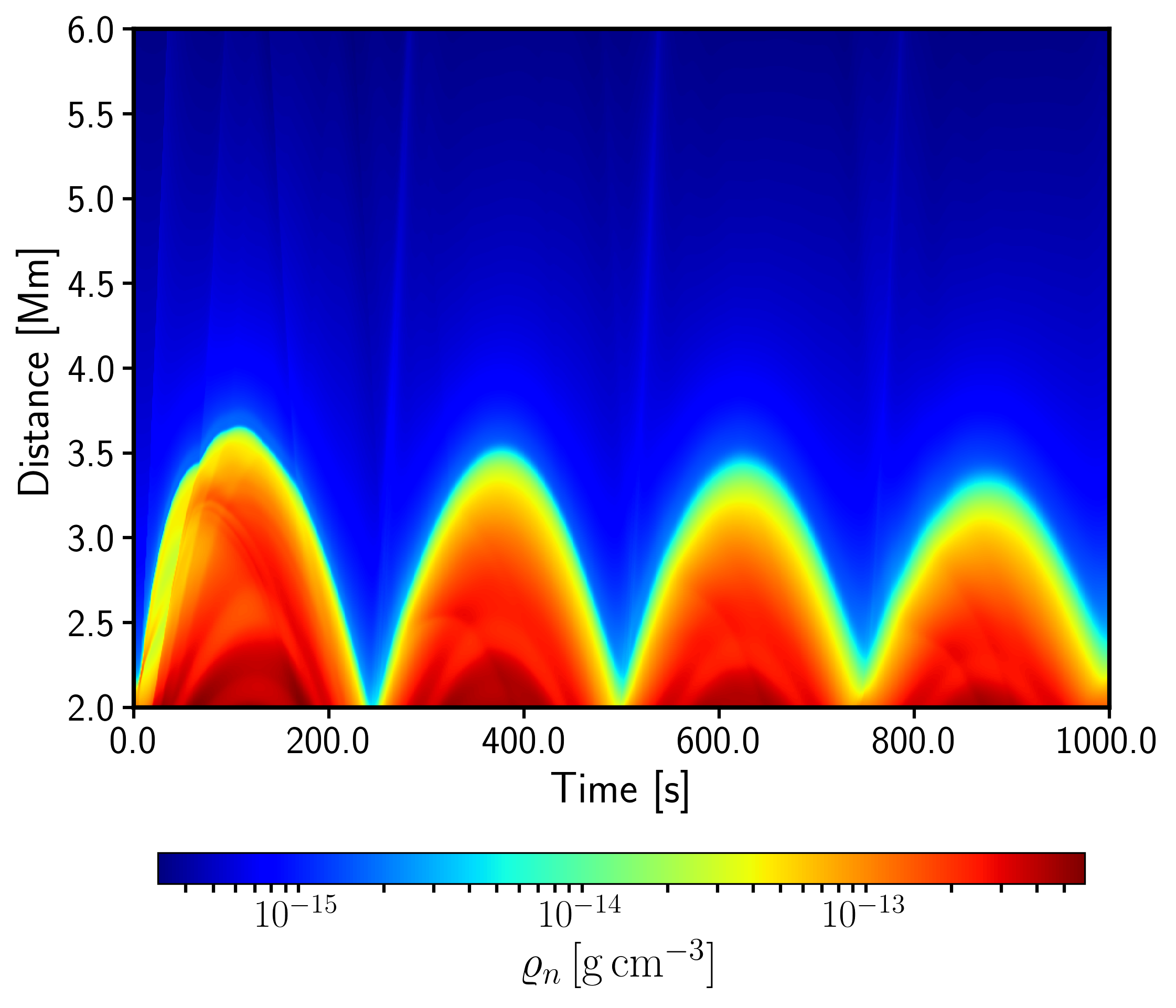}
}
\caption{The maps of the mass density ($\rho$) of ion and neutrals are shown along with the spicular jet in the top panels. The black-dotted slits from 2 to 6 Mm are overlaid on the jet's path along which the distance-time maps are derived and shown in the bottom panels.  These maps reveal a recurrent pattern of spicular jets exhibiting quasi-periodic rise and fall in the model solar atmosphere.}
\label{}
\end{figure*}


\begin{figure*}
\centering

\mbox{
\includegraphics[width=7cm, height=7cm]{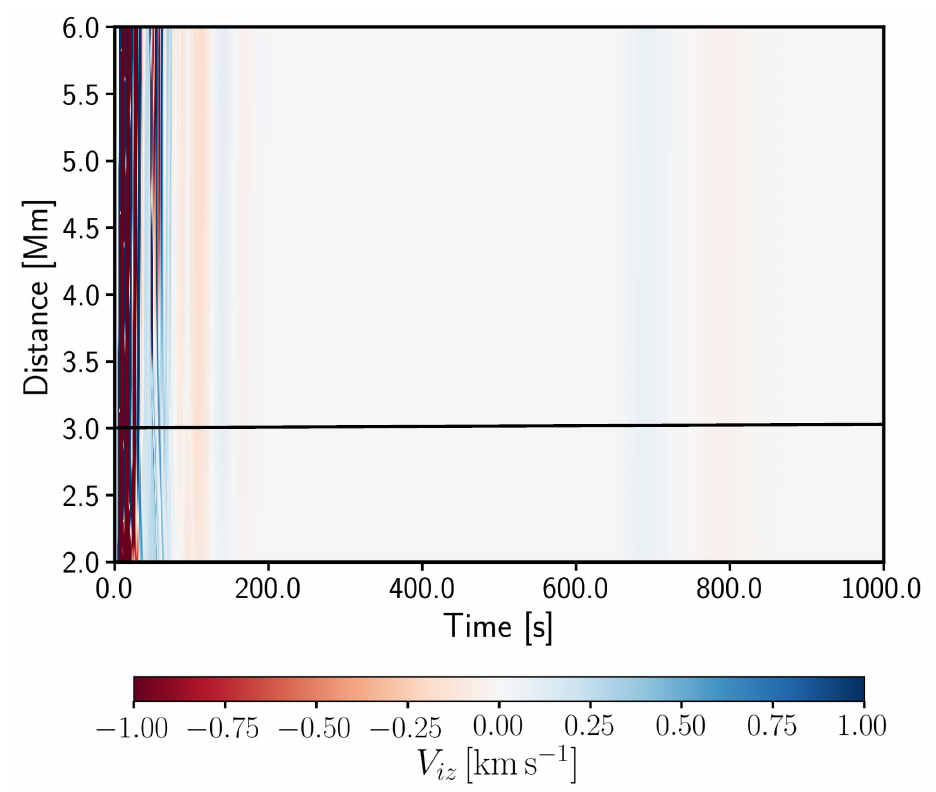}
\includegraphics[width=7cm, height=7cm]{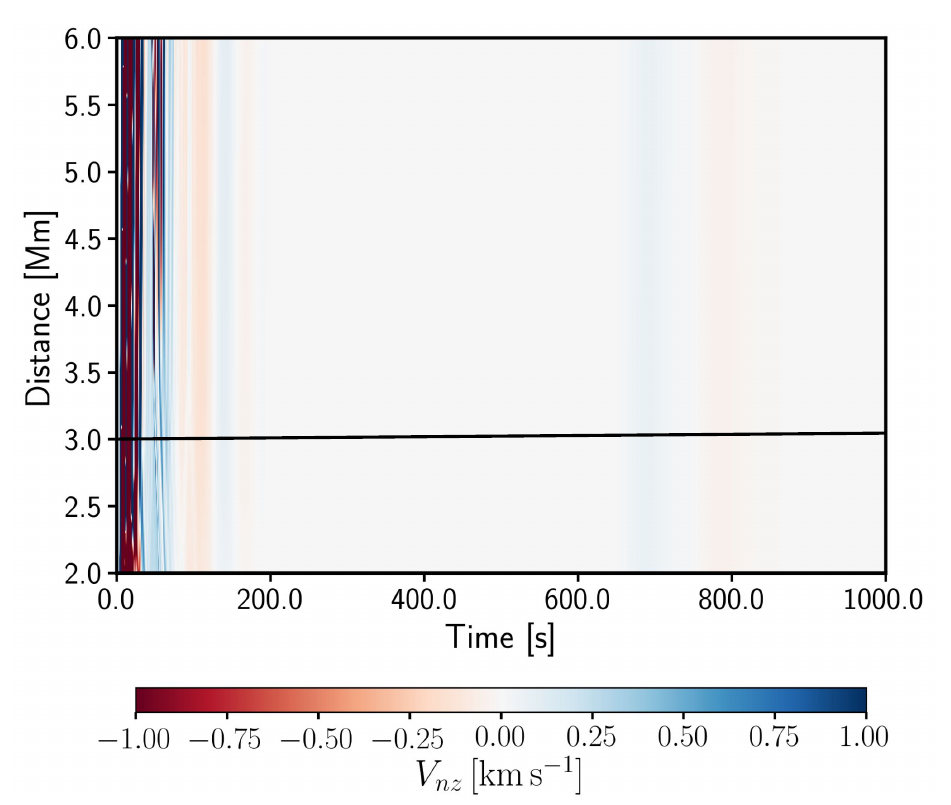}
}
\mbox{
\includegraphics[width=7cm, height=7cm]{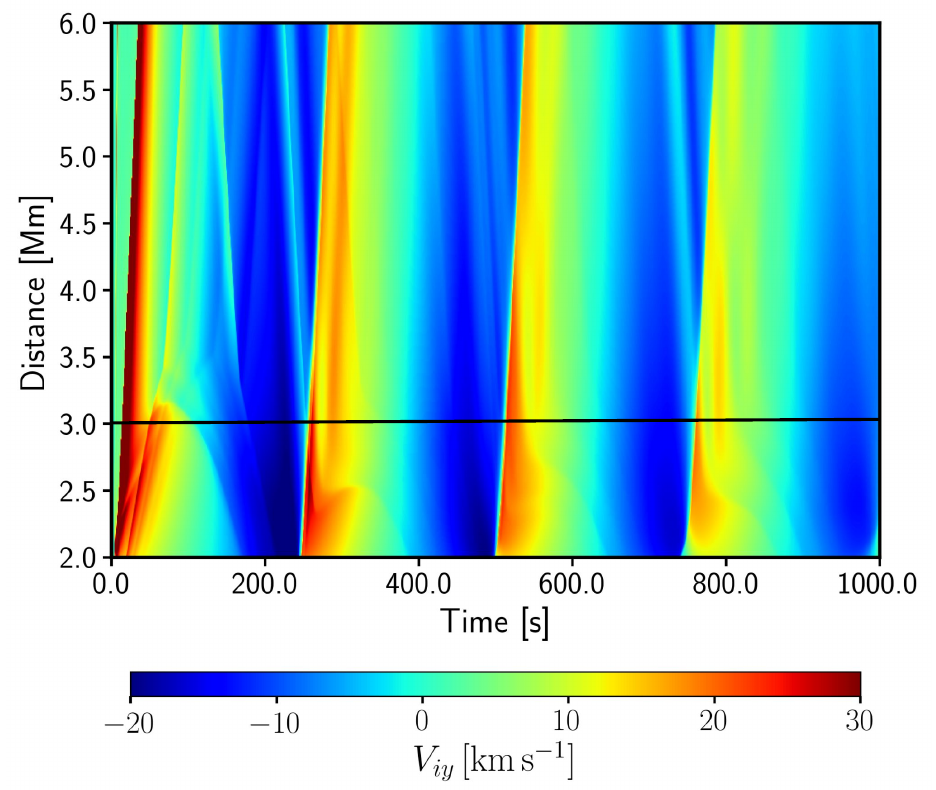}
\includegraphics[width=7cm, height=7cm]{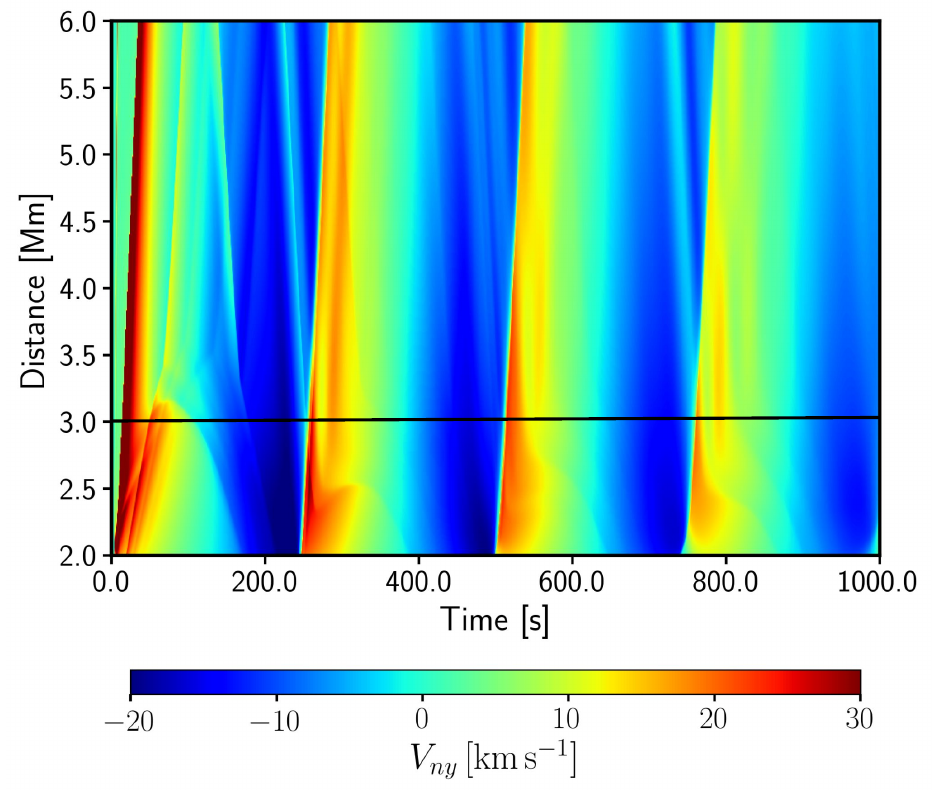}
}

\mbox{
\includegraphics[width=7cm, height=6cm]{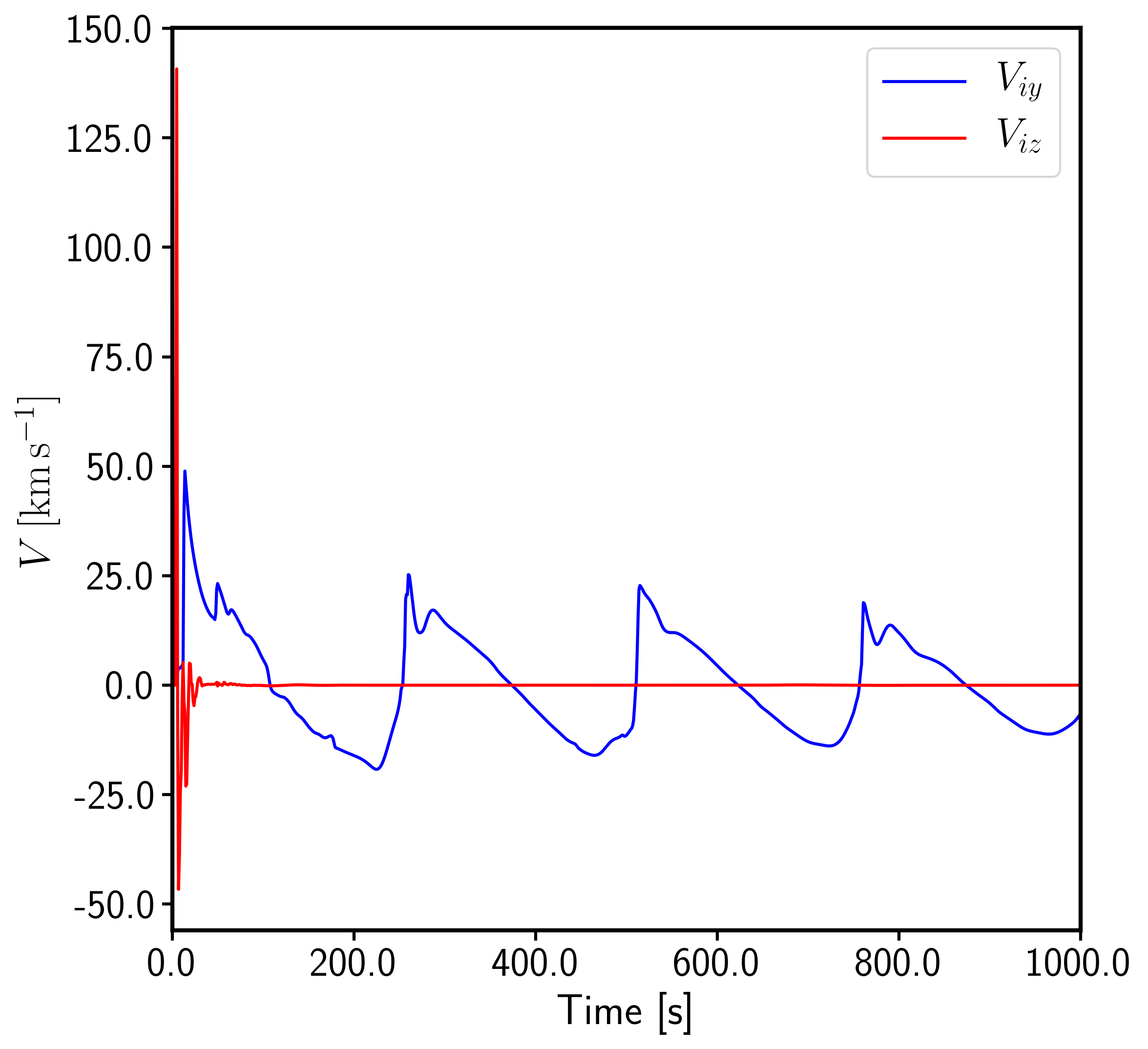}
\includegraphics[width=7cm, height=6cm]{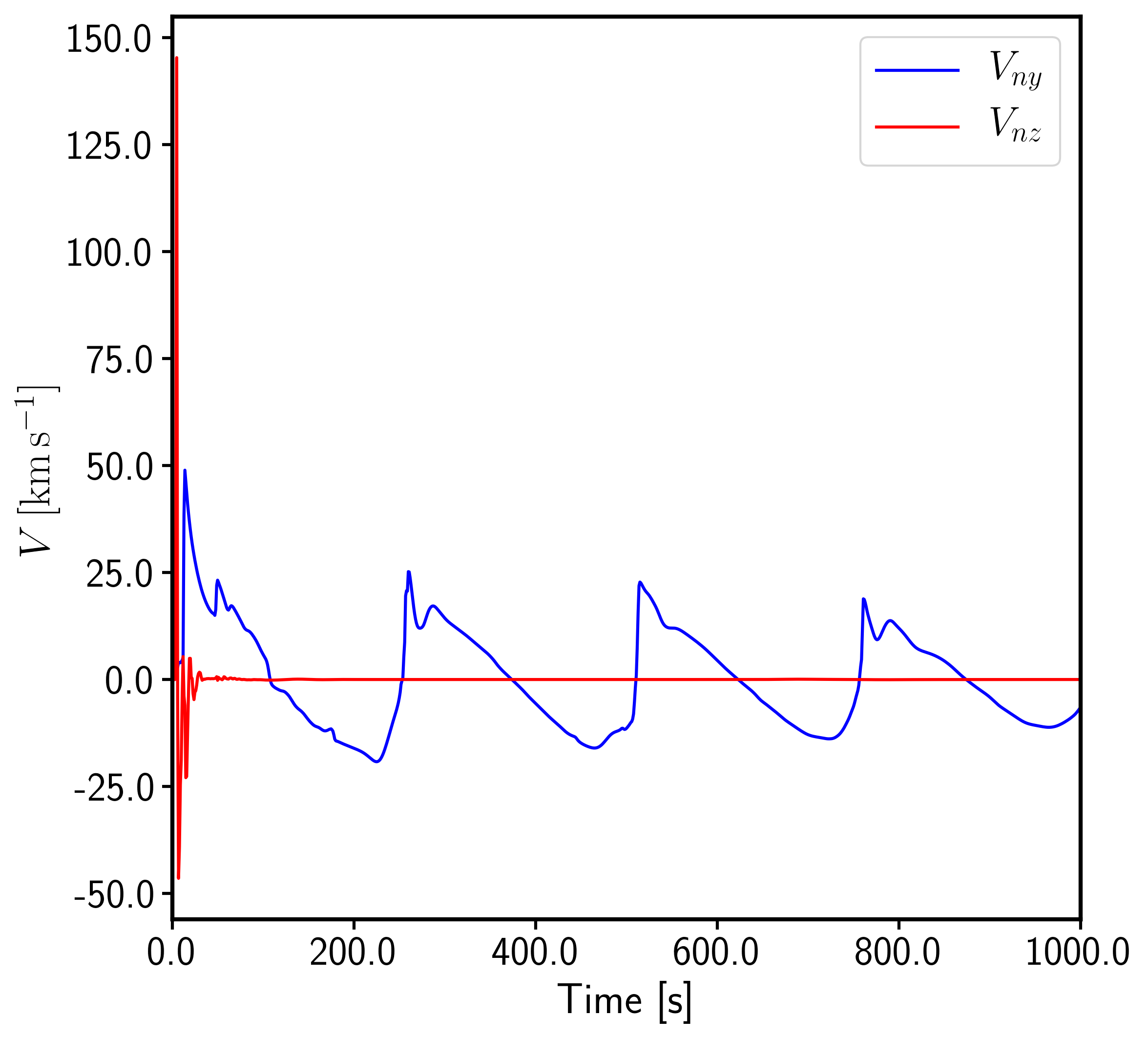}
}

\caption{The distance-time maps of the strength of transverse pulse  {\it akin} to the Alfv\'en pulse (V$_{z}$), the vertical field aligned velocity (V$_{y}$) in ions and neutrals are estimated along the chosen slit (cf., Fig.~4, top-panel). The temporal evolution of V$_{iz}$, V$_{nz}$ (red curve) and V$_{iy}$, V$_{ny}$ (blue curve) along the horizontal black slit taken at y=3.0 Mm (cf., top and middle panels) are shown in the bottom panels. The red curves depict the magnitude of applied Alfv\'en pulse and blue curves indicate the signature of shock propagating above the spicular jets. }
\label{}
\end{figure*}

\section{Results}

We have employed the transverse pulse {\it akin} of Alfv\'en pulse in both ions and neutrals velocities in the lower solar { chromosphere}. The non-linear Alfv\'en pulse is concentrated in a relatively small spatial region of the chromosphere { initially, i.e., 50 km}. Initially, the applied magnitude of the pulse is high enough (150 km s$^{-1}$) to non-linearly coupled with the field-aligned magnetoacoustic perturbations to further drive cool jets along the field lines (Fig.~3). The amplitude of the applied pulse conduct itself as an initial driver in the formation of these cool jets.

Fig.~3 shows the evolution of cool jets in both $\rho _{i}$ and $\rho _{n}$. The velocity vectors (red arrows) overlaid on density maps in Fig.~3 depict the velocity of upward moving and downflowing species in the localised model solar atmosphere. 
The ionization fraction, $\rho_i(y)$/$(\rho_i(y)+\rho_n(y))$, varies in time. { It is} higher in the corona. 
{ It is also suggestive} that due to Lorentz force  magnetic field pushes ions and as they are weakly coupled in the chromosphere with neutrals, later became even less abundant there.  Although this much ion-neutral coupling is sufficient for ions which are driven by $B$ to drag neutrals and create jets seen both in $\rho _{i}$ and $\rho _{n}$ or in their temperatures. { As the cooler chromospheric material has a lower ionisation degree than the corona, 
the ionisation fraction is higher as a result of the spicule. As we do not have ionisation/recombination in the model, which could be heavily active here, our model suffers a drawback in the present simulation in this particular context.
We take a slit along the cool jets (in the ions and neutrals maps as demonstrated in the top panels of Fig.~4) to examine their formation and kinematical properties. We follow these spicular jets by clicking on the path coordinates from 2.0 Mm up to 6.0 Mm, and make the distance-time diagrams in log $\rho_{i,n}$ (see the lower panels of Fig~4). The distance-time maps of the magnitude of the Alfv\'en pulse (V$_{z}$), as well as (V$_{y}$) in ions and neutrals are { calculated} along the same chosen slit and displayed in Fig.~5 top and middle panels.

This cool jet is triggered by the energy transfer from Alfv\'en pulse (cf., Fig.~5, top panel) into the field-aligned perturbation in non-linear regime of two-fluid system. In the 2.5-D two-fluid simulation, this energy transfer { initially} occurs  from applied transverse perturbation into the field-aligned perturbation through ponderomotive force.  Later on, this filed-aligned perturbations generate the magnetoacoustic shocks (Fig.~5, middle panel). Some small differences can be seen on profiles { V$_{i}(y)$ and  V$_{n}(y)$,} but it is not any significant difference (see middle/bottom-left and middle/bottom-right panels in Fig.~5). These shocks are followed by the cool plasma forming the spicular jets\cite{1982ApJ...257..345H,1999JPlPh..62..219V,2013SoPh..288..205T}. These jets are propelled along the magnetic field lines below the shocks in the solar atmosphere. 
The ponderomotive force is evolved in two-fluid system which is responsible as an initial triggering mechanism for the origin of jets rising and falling on the same path (Fig.~4, bottom panel, Fig.~5, first and second rows) which is commonly observed in the chromosphere (Figs.~4, 9)\cite{2004Natur.430..536D, 2022MNRAS.511.4134S}. The initial pulse in the transversal component of ion and neutral velocity excites in a while perturbations in $B_{z}$. The ponderomotive force results from the vertical component of magnetic pressure gradient implied by $B_{z}$. As a result of this force, in a low plasma beta region, essentially vertical ion flow, $V_{iy}$, is generated and the ion-neutral drag,  $V_{iy}-V_{ny}$, becomes increased, as neutrals are not directly affected by magnetic field. These neutrals acquire some momentum from ions until their velocities potentially equalize. As a consequence of the drag, some kinetic energy is thermalized leading to temperature increase, and waves become damped, leading to the reduction of the ponderomotive force. Once the first spicule is triggered under this mechanism, then follow-up jets are the response of downflowing spicular plasma due to an imbalance between the pressure gradient and gravity, as well as rebound shocks. The transport of the energy from Alfv\'en pulse to the field aligned perturbations due to ponderomotive force is considered as an initial trigger mechanism.}
{ Such physical phenomenon was firstly proposed by Hollweg \cite{1982ApJ...257..345H} in 1-D MHD model, who found that the transverse perturbation can produce pure Alfv\'en pulses above solar photosphere hat may couple with magnetoacoustic shocks/wave to launch spicules. }

The similar physical process is modelled and show-cased by \cite{2022MNRAS.511.4134S} in MHD simulations in the ideal plasma. 
In the present work, we study the effect of ion–neutral collisions and thermal conduction in the formation of cool jets in two-fluid regime, in addition to the non-linear transverse perturbation that trigger them. We adopt a two-fluid model to describe the partially ionized plasma, where ions and neutrals interact through collisions. The degree of ionization varies from 0.1\% in the lower chromosphere to 10\% in the upper chromosphere, making the ion–neutral collisions significant for the plasma dynamics. The collisions result in friction and momentum transfer between the two fluids, which modify the propagation and damping of magnetoacoustic shocks/waves. These shocks/waves can heat the localized chromosphere by converting their energy into thermal energy, but they can also be attenuated by the collisions and lose their efficiency. The attenuation of the shocks/waves produces additional heating in the localized region of the chromosphere, which plays a significant role in the formation of spicular jet. 

In addition, we include the effect of thermal conduction for both ions and neutrals in the model solar atmosphere, which { may generically} results in some energy transport towards the solar chromosphere, { thus causing energy re-distribution during the propagation of these cool jets}. { However, this causes some background plasma upflows in the model solar atmosphere that enables more increased kinetic motions of the transition region counterpart of the spicular jets \cite{2023EPJP..138..209S}. In the present model, thermal conduction { may} also create the damping of magnetoacoustic shocks/waves \cite{2023EPJP..138..209S}. 
{ Therefore,} the ion–neutral collisions and thermal conduction may have a substantial impact on the formation, triggering/evolution of chromospheric cool jets by influencing their generation, propagation, dissipation, and thermal properties. The detailed analyses of these candidates on the physical properties of cool jets will be the subject of future study.}

\begin{figure*} 
\centering
\includegraphics[width=14cm, height=9cm]{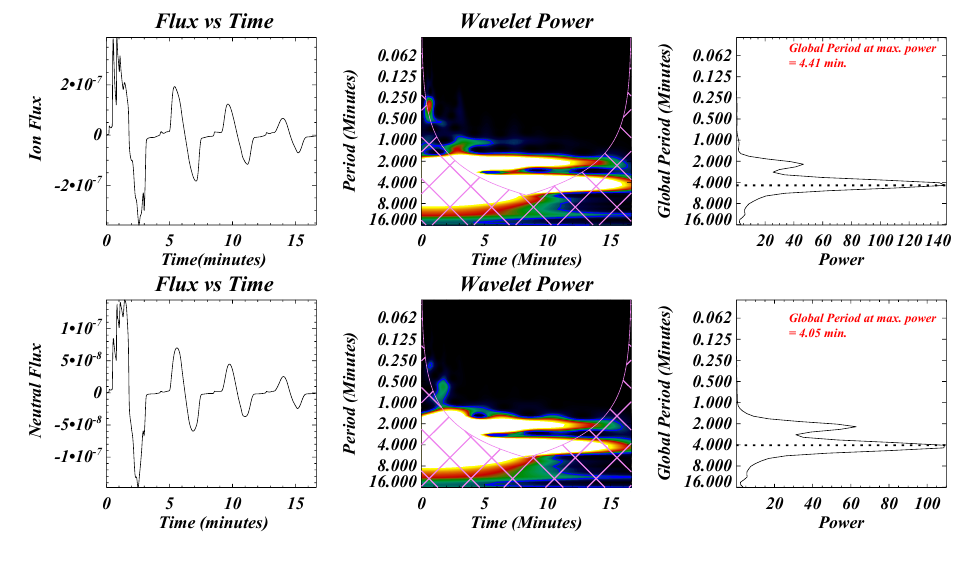}
\caption{The power spectral analyses using wavelet tool is performed on the temporal variation of the mass flux launched by jets for ions and neutrals. The ion and neutral mass fluxes are estimated at 3.0 Mm height just above the TR in the path of spicular jet as a function of time.}
\label{}
\end{figure*}

We perform a wavelet analysis (Fig.~6) on the time-series of estimated mass flux $(\rho v)$ in ions and neutrals { as collected in the overlying corona at $y$=3 Mm} along the path where spicular jets rise and fall (see Fig.~ 6, left column, first and second row) \cite{1998BAMS...79...61T}. The temporal variation of the mass flux (left-column), wavelet spectra (middle-panel) { showing the localized powers in Fourier and time-period space in 2-D space, and global power spectra (right-column; the averaged power over the total time duration for each Fourier period )} are shown in Fig.~6 for ions (top-row)  and neutrals (bottom-row). { By making use of the wavelet analysis tool, the search for the significant oscillatory periods in the given time-series is performed by a time-localized Morlet function that is continuous in both frequency-domain and time \cite{1998BAMS...79...61T}. The wavelet is convolved with the time series (here mass flux of ions/neutrals vs time) to estimate the contributions of the frequencies present in it that resembles accurately with the sinusoidal part by varying scale of the wavelet function. This method deduce the power spectrum of the oscillatory patterns present in the signal (see Fig.~6). Here, we use the wavelet tool to estimate the significant periodicity of the quasi-periodic rise and fall of the mass flux associated with spicular jets.} The mass motions have been calculated with periodicity of $\simeq$ 4.41 min for ions and $\simeq$ 4.05 min for neutrals with a $>$ 95 \% significance level (Fig.~6). Therefore, the mass flux exhibits a quasi-periodic rise and fall of the plasma in the shape of recurrent cool jets with a typical time scale of $\simeq$ 4.0 min.
  

\section{Discussion and Conclusions}

In the  present scientific work, we performed a 2.5-D numerical simulation in two-fluid system (for ions and neutrals) in the presence of thermal conduction and ion-neutral collision to model the triggering of spicular jets caused by Alfv\'en pulse. These jets show the quasi-periodic rise and fall at a typical $\simeq $ 4.0 min time-scale. The present study provides a  numerical modelling of the formation, kinematics, and mass transport due to evolved cool jets. The results show that transverse velocity pulse can produce the strong field-aligned magnetoacoustic perturbations in the solar chromosphere, causing shocks in the stratified overlying atmosphere. This energy exchange can occur in non-linear regime when transverse Alfv\'en-like perturbation is dissipated and the field-aligned motion due to ponderomotive force is originated \cite{1999JPlPh..62..219V, 2015A&A...577A.126M}. Furthermore, these field-aligned perturbations generate the shocks which create lower pressure regions, and drive rise and fall of cool spicular jets quasi-periodically (Fig.~4, bottom panel) to transfer mass into the localized lower corona (Fig.~6). In the present model, we notice that the ion-neutral collisions and thermal conduction may work in the formation of jet and their kinematics. If we exclude these two physical effects from the two fluid simulations, the appropriate height of the jet is not achieved. The jet attains a reasonable height around 4.0 Mm in this two-fluid simulation in the presence of the thermal conduction and ion-neutral collision. However, it should be noted that the driver might have been too strong and further work needs to be done. In the ideal MHD simulations done by  \cite{2022MNRAS.511.4134S}, the cool spicular jets were even attaining suitable height without implementing the dissipative agents like thermal conduction, etc. The presence of non-ideal, two-fluid plasma in the chromosphere is a more realistic physical configuration of the plasma. In such a plasma configuration, the cool spicule like jets are formed during energy re-distribution due to thermal conduction and the presence of ion-neutral collision as seen in the present simulation. Most likely ion-neutral collision dissipate the part of the energy of wave, thus leading to the reduction of the ponderomotive force, even the amplitude of the Alfv\'en pulse was strong initially. Therefore, this force does not lead effective field-aligned perturbations likewise in ideal MHD, and does not intend to launch the spicular like jets to the maximum height until the thermal conduction re-distribute the energy and heating generated upflows support the upward motion of the chromospheric plasma together.

We extend and provided new insights on one previous MHD model given by \cite{2022MNRAS.511.4134S} regarding the formation of spicule-like cool jets by adding thermal conduction and ion-neutral collision effects in two-fluid regime. The previous work is performed in 2.5-D simulation in ideal MHD by implementing multiple Alfv\'en pulses at the solar chromosphere \cite{2022MNRAS.511.4134S}. In that model, randomly generated multiple Alfv\'en pulses of magnitude between 50-90 Km s$^{-1}$ were applied in the solar chromosphere between 1.5-2.0 Mm. It was demonstrated that the Alfv\'en pulses generate field-aligned magnetoacoustic perturbations that evolve into shocks, which drive spicule-like mass motions. This mechanism was first proposed by Hollweg \cite{1982ApJ...257..345H} in 1-D MHD model, who argued that the transverse perturbation can produce pure Alfv\'en pulses that may couple with magnetoacoustic shocks/wave responsible for the generation of spicule-like jets. 

We performed the similar simulation set-up as reported by \cite{2022MNRAS.511.4134S}, however, in two-fluid regime of the solar plasma by also invoking the effects of ion-neutral collisions and thermal conductivity. Our previous numerical modeling by Singh et al. \cite{2022MNRAS.511.4134S} describe the generation of spicule-like cool jets in a similar fashion using ideal  and single fluid, 2.5 D MHD. Therefore, our aim here is to just show-case that physical model of the cool jets under more realistic physical conditions, i.e.,  two fluid regimes including thermal condition and ion-neutral collisions. We do not focus on the parametric study in the present case. 
Its interesting to note that in the presence of the more appropriate physical conditions of non-ideal partially ionized plasma the spicular jets  attain suitable height and appear as a thin plasma column rising and falling quasi-periodically as seen in various observations. When we switched on the thermal conductivity  the re-distribution of energy and the damping of waves lead to the formation of these spicule-like cool jets. We assert finally that to understand the exact physics of cool chromosperic jets and their formation, the two-fluid simulations with ion-neutral collision and thermal conductivity { may} be taken into account in numerical models. {
Additional physical processes, such as radiative transfer, radiative cooling, radiative transport, ionization, recombination, and heating terms, should be incorporated into future models. These additions will enhance the comprehensiveness and accuracy of the simulations \cite{2023MNRAS.525.4717S,2021ApJ...911..119Z}. MHD may also be an appropriate { formalism} but it rules-out neutrals that are present in the lower solar atmosphere and might be at work there in formation of spicular jets.

\ack{{ We thank the two reviewers for their critical and valuable scientific remarks that improved the manuscript considerably.} A.K.S. acknowledges the ISRO Project Grant number DS\textunderscore2B-13-12(2)/26/2022-Scc2 for the support of his research. BS gratefully acknowledges the Aditya-L1 Support cell, which is a joint effort of Indian Space Research Organization (ISRO) and Aryabhatta Research Institute of Observational Sciences (ARIES), Nainital, for support of current research and providing him with a postdoctoral research grant. K.M.'s work was done within the framework of the project from the Polish Science Center (NCN) Grant No. 2020/37/B/ST9/00184.  BS \& AS also acknowledge the National Supercomputing Mission (NSM) for providing computing resources for ‘PARAM Shivay’ at the Indian Institute of Technology (BHU), Varanasi, which is implemented by C-DAC and supported by the Ministry of Electronics and Information Technology (MeitY) and Department of Science and Technology (DST), Government of India. AS thanks the financial support from IIT (BHU) to IDD/M.Tech students for academic works. The authors also thank the developers of the JOANNA two-fluid code and the PYTHON libraries for their useful tools for numerical data analysis. We also thank Ms Kartika Sangal for her assistance with the wavelet analysis. We acknowledge the use of JOANNA code developed by D. Wojcik. D.Y. is supported by the National Natural Science Foundation of China (NSFC; grant numbers 12173012, 12111530078 and 11803005), the Guangdong Natural Science Funds for Distinguished Young Scholar (grant number 2023B1515020049), the Shenzhen Technology Project (grant number GXWD20201230155427003-20200804151658001) and the Shenzhen Key Laboratory Launching Project (grant number ZDSYS20210702140800001). T.V.Z. was supported by the Shota Rustaveli National Science Foundation of Georgia (project  FR-23-6815-2).}


\appendix

\section{Data Availability}
The numerical simulation data that underpins this research is accessible at Dryad (https://doi.org/10.) 1/dryad.jm63xsjjs) and can be shared with the corresponding author upon reasonable request.




\bibliographystyle{vancouver}
\bibliography{sample}

\end{document}